\documentclass[12pt]{iopart}
\usepackage{graphicx}
\usepackage{epstopdf}

\begin{document}
\title{An introduction to stochastic self-assembly: theory, simulation and experimental applications}

\author{E. M. Schwen}

\address{Washington and Lee University, 204 W. Washington Street, Lexington, VA 24450, USA}

\ead{schwene15@mail.wlu.edu}

\begin{abstract}

We introduce three stochastic cooperative models for particle deposition and evaporation relevant to ionic self-assembly of nanoparticles with applications in surface fabrication and nanomedicine. We present a method for mapping a stochastic model onto the Ising model, which allows us to use the established results for the Ising model to describe the properties of the system. After completing the mapping process, we investigate the time dependence of particle density using the mean field approximation. We complement this theoretical analysis with Monte Carlo simulations that support our models. These techniques, which can be used separately or in combination, are useful as pedagogical tools because they are tractable mathematically and they apply equally well to many other physical systems with nearest-neighbor interactions including voter and epidemic models.

\end{abstract}

\section{Introduction}

Complexity is a new interdisciplinary field of science with the goal of studying how the interactions between components of a system give rise to its collective behavior. Centers for complexity studies are becoming increasingly common. Interest in the field is also evident in the physics curriculum, where the number of courses in modeling complexity is increasing at both the graduate and undergraduate levels. In these courses students use statistical physics methods and computer simulations to study a variety of complex systems.


Analytical and computational  models have proven successful in describing diverse physical systems ranging from surface deposition and chemisorption on crystal surfaces \cite{liggett} to epidemic problems \cite{bio2,bio} and voting behavior \cite{castellano}.  The dynamics of nanoparticle deposition is another application of such models.  It is currently an  active area of research in nanotechnology studies \cite{heflin} which addresses interesting open questions on the theoretical front \cite{gouet}.

Two classes of models that have been particularly successful are \it random sequential adsorption \rm  (RSA) \cite{evans, cadilhe, cadilhe1}, in which particles are adsorbed/deposited at a fixed rate at random unoccupied sites on a grid, and \it cooperative sequential adsorption \rm (CSA) \cite{privman}, in which adsorption rates depend upon the occupation of neighboring sites. One-dimensional sequential adsorption models have been studied thoroughly in different physical contexts \cite{evans, privman}, but adsorption in two dimensions is not as clearly understood. There are many computational adsorption models \cite{redner}, but few analytical solutions have been developed for the general two-dimensional case. Recently, analytical results have been reported for the random sequential process \cite{cadilhe} and reaction-diffusion processes on Cayley trees and Bethe lattices \cite{ben avraham, matin, ali, JSTAT,tome,fonseca}. Adding  the possibility of particle detachment, or \textit{evaporation},  to such models brings additional complications.  One of the standard tools used to study these systems, the \emph{empty-interval method} \cite{redner}, fails when evaporation is considered. Evaporation has been treated analytically in a few studies of one-dimensional systems using a quantum mechanical approach \cite{grynberg}.

In this thesis, we present a few example stochastic models and use them to introduce the basic techniques employed to study complex systems and the methods used to solve them. This thesis provides a broad overview of the recent work completed by our research group \cite{concentration paper, EJP, Linear Model, Total Lattice Model, Trees Paper}.

The first two models we introduce are general stochastic models for particle deposition and evaporation where the deposition rate at each site is determined by the occupation of the neighboring sites. The third stochastic model also addresses particle deposition and evaporation, but the deposition rate is instead determined by the occupation of all sites on the lattice. While we present each model in terms of particle deposition and evaporation, they can be easily modified to describe any system governed by nearest-neighbor interactions for the first two models or lattice-wide interactions for the third model. We introduce these models as templates for a large class of cooperative stochastic models. We present specific applications of our models to ionic self-assembled thin films and encapsulation of drug molecules, but we also invite readers to apply these and similar models to other suitable physical systems.

The main analytical methods we use to investigate our stochastic models are comparison with the Ising model and the mean field approximation. The Ising model \cite{Ising}, originally developed to explain ferromagnetism, is one of the best-known stochastic models and provides an excellent framework to study a variety of other physical systems. It considers both nearest-neighbor interactions between spins and the the effect of an external field on each individual spin. The Ising model is one the most studied and versatile models in equilibrium statistical physics, well-covered in statistical physics courses and literature \cite{redner}. The mean field approximation is an efficient analytical tool used to study cooperative systems with relatively uniform fields throughout.

Computer simulations utilizing Monte Carlo methods allow further investigations into the dynamics of particle deposition and related systems. We outline a basic algorithm, useful as a learning tool, and an event-driven algorithm that can be effectively implemented on a larger scale. We present simulation results for both square lattice and Cayley tree geometries throughout the thesis and outline possible research applications for computational studies.

We begin by presenting our cooperative power model. This model is the most general of the three and we will use to introduce the analytical, computational, and experimental techniques used throughout the thesis. We present a method for mapping the model onto the Ising model (defined on a square lattice or a Cayley tree). Using the detailed balance condition, we relate the deposition and evaporation rates in our model to the Ising model coupling and external field constants. We then use  known Ising model results  to describe the steady-state properties of our model. We use the mean field approximation to study the time dependence of the particle density, presenting the basic assumptions and techniques involved and employing them to derive the rate equation for the particle density.

We then move to specific applications of our cooperative power model to ionic self-assembly of nanoparticles and drug encapsulation using dendrimers. The ionic self-assembly section applies our model specifically to a two-dimensional lattice geometry. We describe our experimental technique for creating thin films and compare our analytical solutions with both computer simulations and experimental data. The drug encapsulation section applies our cooperative power model to a Cayley tree geometry and presents analytical solutions for particle density by generation alongside simulation results.

After the investigation of our cooperative power model, we move on to address the cooperative linear model. The cooperative linear model uses a linear approach to calculate the effects of the nearest neighbors but is quite similar to the cooperative power model in both theoretical methods and experimental applications. We introduce the model, investigate it using Ising model mapping and the mean field approximation, and compare the analytical results with both computer simulations and our ionic self-assembly experimental data.

The final model we present is our total lattice cooperative model. Since this model calculates the cooperative effect based on the total occupation of the lattice, some of the mathematical tools and results are quite different from the other two models. We first present a rate equation and general solution for the particle density using the mean field approximation. We then use a master equation approach and matrix theory to calculate a general time-dependent solution for the probability for any number of cells to be filled. We compare the model results to both computer simulations and ionic self-assembly experimental data.

Our thesis is structured as follows: We begin in section 2 with a presentation and analysis of our cooperative  power model. In sections 2.1 and 2.2, we map our cooperative power model onto the Ising model and derive a rate equation for the particle density using the mean field approximation. In section 2.3, we discuss Monte Carlo simulations and outline two general algorithms for simulating the model dynamics. In section 2.4, we present the applications of our cooperative power model to ionic self-assembly of nanoparticles. In sections 2.5, we apply our cooperative power model to the process of drug encapsulation on a Cayley tree geometry. We introduce and analyze our cooperative linear model in section 3. In section 4, we present our total lattice cooperative model and results. We conclude in section 5 with a summary of our thesis and an analysis of further uses of our models and methods.

\section{Cooperative power model}

Adsorption kinetics is a very active area of research, particularly in nanotechnology studies. Analytical sequential adsorption models have proven successful in one dimension for modeling surface deposition, polymer chain dynamics, chemisorption on crystal surfaces, epidemics problems, and voting behavior \cite{evans, privman, castellano, bio, bio2, liggett}. 

We present here a general  model that includes both  attachment  and evaporation  of particles; we refer to it as a \it cooperative sequential adsorption with evaporation \rm (CSAE) model. The generic term ``particle" in this model can refer a drug molecule for a drug encapsulation process, a silica nanoparticle for the creation of thin films, or it can be used as an abstract representation for a voting pattern in a voter model or infected individual in an epidemic model. 

The model is defined on a discrete lattice of $N$ sites that can be either empty or occupied. The occupation of site $i$ is defined by an  an occupation number $n_{i}$:  $n_{i}=1$ if occupied, $n_{i}=0$ if empty. Particles attach at site $i$ with a probability rate equal to $\alpha \beta^{\eta}$, where $\eta=\sum_{j \in NN}n_{j}$ is the sum over all occupied nearest-neighbors of site $i$. This rate mimics electrostatic interactions and space constraints through its dependence on $\eta$. For the moment, we restrict  $\beta$ to values less than 1.  The same model, for  $\beta$ larger than 1, can be used to describe epidemic or voter models, where the presence of nearest-neighbors increases the chances for a site to change its state. The model also considers evaporation of particles with rate $\gamma$, independent of the occupation of neighboring sites. 

The change from an empty to a filled state  of a site of occupation $n_{i}$ and vice versa can be described by the transition rate:
\begin{equation}
c(n_{i}\rightarrow(1-n_{i}))=\gamma n_{i}+(1-n_{i})\alpha\beta^{\eta}.
\end{equation}

This model serves as a template for a wide variety of models, tailored to specific processes.

\section*{2.1. Mapping onto the Ising model}

The theoretical framework of equilibrium statistical mechanics \cite{gibbs} establishes that for a system in contact with a heat reservoir of temperature $T$, the probability of finding the system in configuration $s$ is given by the canonical distribution:
\begin{equation}
P_{eq}(s)=\frac{e^{- H(s)/kT}}{Z},
\end{equation}
where $k$ is Boltzmann's constant, $H(s)$ is the microscopic Hamiltonian, and the partition function $Z$ ensures the normalization of the probability. Thus, once we have labeled the microscopic configurations, $s=\{s_i\}$, and  have determined the Hamiltonian, $H(s)$, we can in principle calculate the partition function of the equilibrium system and average values of time-independent observables.

The Ising model was introduced by Heinrich Lenz \cite{lenz} in 1920 to understand the nature of phase transitions in ferromagnets and was solved by Ernst Ising \cite{Ising} in 1925. Over the years, the Ising model proved to be extremely versatile and was used to describe systems as diverse as the spread of rumors and the phase transitions of water. It is amenable to both computer simulation studies and analytical solutions. 

In general, the Hamiltonian associated with the  $d$-dimensional Ising model of a system of N spins ($s_{i}$, $1 \le i \le$ N) in an external field is:

\begin{equation}
H=-J\sum_{i,j\in NN}s_{i}s_{j}-B\sum_{i=1}^{N}s_{i}.
\end{equation}

The first sum is over all distinct nearest-neighbor spin pairs, and the second sum accounts for the interaction of each individual spin with the external field. $J$ is known as a \it coupling constant \rm between spins, and $B$ represents the external field. The spin numbers  are defined as $s_{i}=1$ (spin up) and $s_{i}=-1$ (spin down). The equilibrium properties of this model can be derived from the partition function.  In order to study time-dependent behavior of an Ising-type system, however, a different approach is needed.

In a seminal paper, Glauber \cite{glauber} answered this challenge by solving the kinetics of a one-dimensional Ising spin model in an external magnetic field. The starting point is the \it master equation\rm,  which expresses the conservation of configurational probabilities:

\begin{equation}
\frac{dP(s,t)}{dt}=\sum_{s'}\{c(s'\rightarrow s)P(s',t)-c(s\rightarrow s')P(s,t)\}.
\end{equation}

The time-dependent probability $P(s,t)$ of finding the system in configuration $s$ at time $t$ changes due to the transfer of probability into $s$ from other configurations (a \it gain \rm term), or from $s$ into others (a \it loss \rm term). The evolution of $P(s,t)$ is dictated by a set of transition rates $c(s\rightarrow s')$ from configuration $s$ to $s'$. For a spin system, one configuration leads into another via a spin flip. 

To solve the steady-state problem we need to find the stationary solution of the above equation:
\begin{equation}
0=\sum_{s'}\{c(s'\rightarrow s)P_{eq}(s')-c(s\rightarrow s')P_{eq}(s)\}.
\end{equation}

Glauber showed that the transition rates must be chosen so they  satisfy the detailed balance condition:
\begin{equation}
c(s'\rightarrow s)P_{eq}(s')=c(s\rightarrow s')P_{eq}(s).
\end{equation}
This expresses conservation of probability currents (in both directions) between all possible  configuration pairs.

Inserting the equilibrium probability distribution, Eq. (2), into the detailed balance condition gives:

\begin{equation}
\frac{c(s'\rightarrow s)}{c(s\rightarrow s')}=e^{\Delta H/kT}
\end{equation}
where $\Delta H=H(s')-H(s)$ is the change in the energy of the system when one spin is flipped.

To relate our model to the Ising model, we write our particle transition rates from Eq. (1) in terms of spin numbers by associating a \it spin up \rm with a filled site  and a \it spin down \rm with an empty site, thus:
\begin{equation}
 n_{i}=\frac{1+s_{i}}{2}.
\end{equation}
The transition rate becomes: 

\begin{equation}
c(s_i)=\frac{1+s_{i}}{2}\gamma+\frac{1-s_{i}}{2}\alpha\beta^{\sum_{j \in NN}\frac{1+s_{j}}{2}}.
\end{equation}
Defining:
\begin{eqnarray}
K\equiv\frac{J}{kT}\\
h\equiv\frac{B}{kT},
\end{eqnarray}
the detailed balance condition becomes (after simplifications):
\begin{equation}
\frac{c(s)}{c(s')}=\frac{P_{eq}(s')}{P_{eq}(s)}=\frac{e^{-Ks_{i}\sum_{NN}s_{j}-hs_{i}}}{e^{Ks_{i}\sum_{NN}s_{j}+hs_{i}}},
\end{equation}
where $s$ denotes any of the $2^{N}$ possible configurations of all $N$ spins in the system, and $s'$ is the state with one spin flipped. $K$ and $h$ can be found from the detailed balance condition. 

To exemplify the method, we pick a one-dimensional case for which the number of neighbors of spin $s_{i}$ is 2. The rate $c$ becomes:

\begin{equation}
c(s_i)=\frac{1+s_{i}}{2}\gamma+\frac{1-s_{i}}{2}\alpha\beta^{(1+\frac{s_{i-1}+s_{i+1}}{2})},
\end{equation}
and the detailed balance condition becomes:
\begin{equation}
\frac{\frac{1+s_{i}}{2}\gamma+\frac{1-s_{i}}{2}\alpha\beta^{(1+\frac{s_{i-1}+s_{i+1}}{2})}}{\frac{1-s_{i}}{2}\gamma+\frac{1+s_{i}}{2}\alpha\beta^{(1+\frac{s_{i-1}+s_{i+1}}{2})}}=\frac{e^{-Ks_{i}\sum_{NN}s_{j}-hs_{i}}}{e^{Ks_{i}\sum_{NN}s_{j}+hs_{i}}}.
\end{equation}
This holds for each of the eight distinct cases of the set ($s_{i}$, $s_{i+1}$, $s_{i-1}$), providing enough independent equations to determine:
\begin{eqnarray}
K=\frac{1}{4}ln(\beta)\\
h=\frac{1}{2}ln(\frac{\alpha\beta}{\gamma}).
\end{eqnarray}

The general case, where each site has $z$ nearest-neighbors, yields:
\begin{eqnarray}
K=\frac{1}{4}ln(\beta)\\
h=\frac{1}{4}ln(\frac{\alpha^2\beta^z}{\gamma^2}).
\end{eqnarray}

These identifications allow application of  well-established Ising model results to different types of lattices. For our model, we can calculate the particle density of the steady state from the magnetization of corresponding  Ising spin system, $M=<s_i>$. The relationship between particle density $\rho$ and magnetization is given by: $\rho=(1+M)/2$. The Ising model magnetization of a spin system in an external magnetic field has analytical solutions for one- and two-dimensional lattices, as well as for Cayley trees \cite{baxter}.

\section*{2.2. Rate equation for the particle density and mean field approximation}

While the steady-state properties of the system are easily found from the Ising model results, the time dependence of the particle density is much more challenging. We can derive a rate equation for the overall particle density more easily using the mean field approximation. This replaces the local ``field" created by nearest-neighbor interaction with a mean field that is averaged over the entire lattice \cite{redner}. The approximation assumes that the distribution of particles is essentially uniform; each site will have roughly the same number of occupied nearest-neighbors and therefore will feel the same effect. Mathematically, this means that neighboring sites are uncorrelated. The ensemble average of the nearest-neighbor correlations is therefore approximated by the product of the mean individual site occupations:

\begin{equation}
\langle n_in_j\rangle=\langle n_i\rangle\langle n_j\rangle.
\end{equation}

Using this approximation, we arrive at the following equation for the time-dependent mean site occupation $i$:
\begin{equation}
\frac{\partial \langle n_i\rangle}{\partial t}=-\gamma \langle n_i\rangle+(1-\langle n_i\rangle)\alpha\beta^{\langle\eta\rangle},\\
\end{equation}
with 
$\langle \eta\rangle=\sum_{j\in NN}\langle n_j\rangle$.
This rate equation describes the change with time of the average occupation number of site $i$. It has a loss term, which represents the possible evaporation of a particle at site $i$ with rate $\gamma$, and a gain term due to the deposition of a particle if site $i$ is empty. The deposition rate depends on the number of occupied neighbors to incorporate the cooperative effects.

Technically, one has to write such rate equations for all $N$ sites of the system. To further simplify the problem, we assume translational invariance, which allows us to remove the location dependence from the site averages:
\begin{eqnarray}
 \langle n_i\rangle  =\langle n\rangle \\
\langle\eta\rangle  	=z\langle n\rangle,
\end{eqnarray}
where $z$ is the number of nearest-neighbors for each site. This approximation is  reasonable for systems where coverage is essentially uniform and edge effects are negligible.

The particle density is defined as:
\begin{equation}
 \rho=\frac{\sum \langle n_i\rangle}{N},
\end{equation}
where $N$ is the total number of lattice sites, leading to a rate equation for the particle density:

\begin{equation}
\frac{\partial \rho}{\partial t}=-\gamma \rho+(1-\rho)\alpha\beta^{z\rho}.
\end{equation}

For the steady state, $\frac{\partial \rho}{\partial t}=0$, this is a transcendental equation:

\begin{equation}
\rho=\frac{\alpha\beta^{z\rho}}{\gamma+\alpha\beta^{z\rho}}.
\end{equation}

Although Eq. (25)
is a nonlinear function, a linear approximation matches the numerical solution well, which is also shown in Fig. 6. We linearize Eq. (25) by performing a Taylor expansion about $\beta=1$:  

\begin{equation}
\rho=\rho(\beta=1)+(\beta-1)\left.\frac{\partial\rho}{\partial\beta}\right|_{\beta=1}+\dots
\end{equation}
where we only keep the linear term.
Using $\rho(\beta=1)=\frac{\alpha}{\gamma+\alpha}$, we obtain the following result:
\begin{equation}\label{theory-linear}
\rho=\frac{\alpha}{\gamma+\alpha}-(1-\beta)\left[4\left(\frac{\alpha}{\gamma+\alpha}\right)^2\left(1-\frac{\alpha}{\alpha+\gamma}\right)\right].
\end{equation}

 Eq. (24) and Eq. (25) can each also be solved numerically using standard software such as Maple or Mathematica. In Fig. 1, we present the numerical solutions for time dependent particle density for three different values of $\beta$ (0.1, 0.5, and 0.9) at a fixed $\gamma=0.3$ and $\alpha=1$ as predicted by Eq (24).

\begin{figure}[h]
\centering
\includegraphics[width=4.5in,height=3.00in,keepaspectratio]{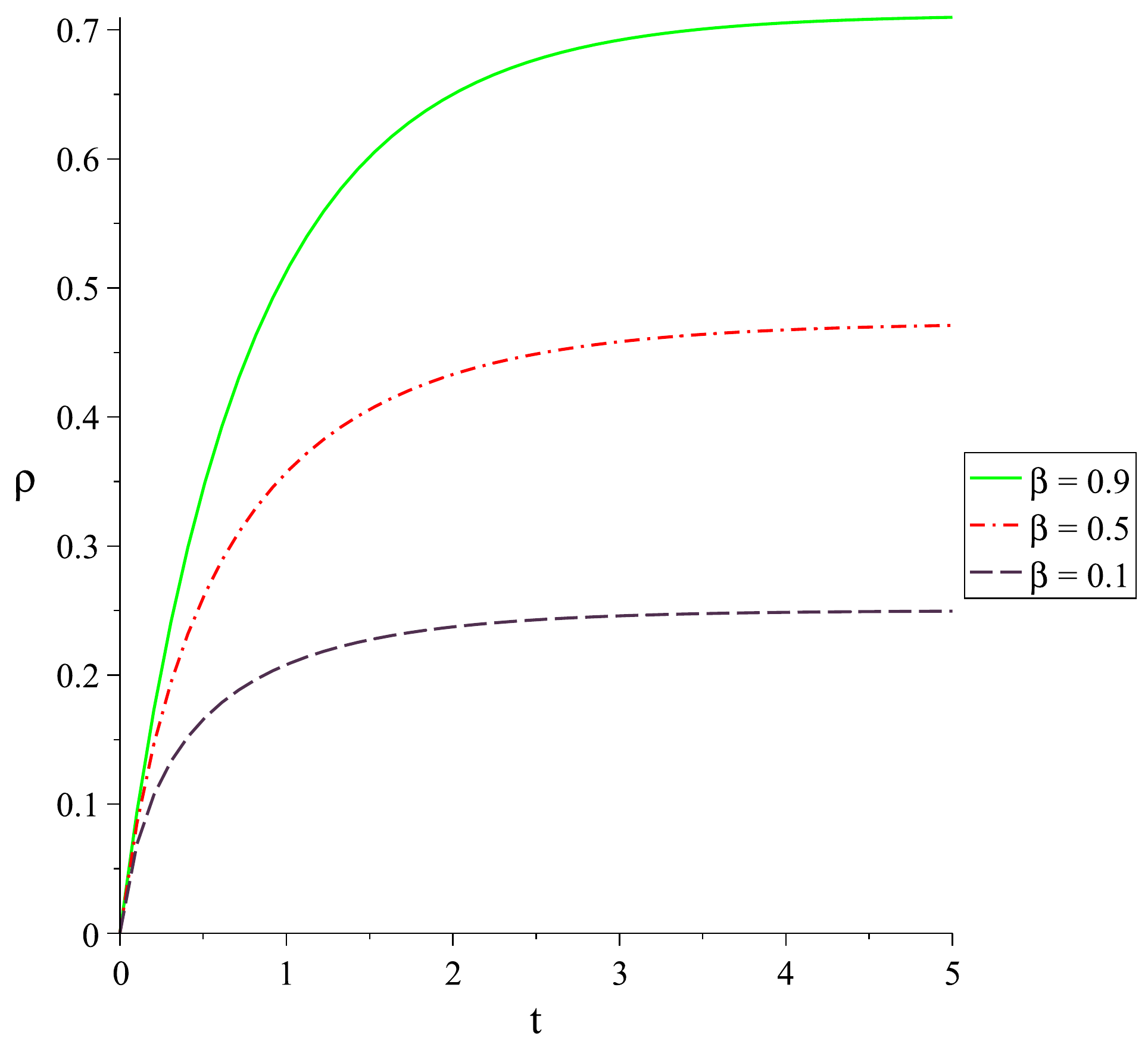}
\caption{Particle density as a function of time for three values of $\beta$ with $\gamma=0.3$ and $\alpha=1$. The three values of $\beta$ shown are (black, dashed line) 0.1, (red, dot-dashed line) 0.5, and (green, solid line) 0.9.}
\end{figure}

For all three $\beta$'s the shape of these curves is the same, but the steady state values of the particle density are drastically different. The values chosen with $\beta < 1$ reflect a physical situation with repulsion between particles. The case of $\beta>1$ corresponds to a physical situation in which the presence of occupied neighbors favors adsorption. This choice can apply to a voter- or an epidemic-type  model. The numerical solutions for  $\beta>1$  are very similar in shape to the ones  presented in Fig. 1 for $\beta<1$. The main difference is the increased rate at which the lattice fills up to $100\%$ coverage. The cooperative aspects of the model disappear for $\beta=1$; it becomes a Langmuir model \cite{langmuir}.

It is interesting to see that Eq. (25), derived independently using rate equations with no relation to the Ising spin model, can be shown to be  identical to the equation derived for the Ising model magnetization in the mean field approximation \cite{redner}:
\begin{equation}
M=\tanh(4KM+h),
\end{equation}
with  $\rho=\frac{1+M}{2}$ and the coupling constants $K$ and $h$ given in Eqns. (17) and (18).

\section*{2.3. Monte Carlo Simulations}

Computer simulations can provide an excellent complement to theoretical analysis for particle deposition and other related models. While analytical results can concisely describe the kinetics and steady state of the system, final solutions are often approximations and can be limited to specific cases or conditions. Computer simulations can provide validation for the theoretical methods when applied to the same conditions, or can be used to extend the analysis to more complex configurations that cannot be addressed theoretically.

Particle deposition models are often simulated using Monte Carlo algorithms. These utilize an intuitive serial process that can be easily programmed by any student with computer programming experience and can serve as an excellent introduction to computational research. The first step for any Monte Carlo simulation is to define the lattice structure. The basic simulation then proceeds as outlined below \cite{redner, gould2}: 

1. Pick a random site (site $x$).

2. Check the occupation of site $x$ and its nearest-neighbors and calculate the probability $p$ of site $x$ changing its occupation state.

3. Generate a random number $0\le r \le 1$ and update the occupation of site $x$ if $r \le p$.

4. Advance the time: $t \rightarrow t+\Delta t$.

These steps are repeated for a specified  time, usually chosen large enough to ensure a steady state has been reached. As the system approaches an equilibrium, however, this simplistic algorithm can become extremely inefficient. Consider a system with much faster deposition than evaporation. As the system reaches nearly full coverage, it becomes increasingly likely that the simulation will select an occupied site and take no action. 

This inefficiency can be eliminated by using an event-driven algorithm where each time step is a guaranteed site update. This method can be implemented with a modified site selection process: instead of using completely random site selection, we calculate the relative probability of all possible site updates at each time step and then randomly select one update event. To implement this modified procedure, we first divide the sites into categories based on their transition rate: occupied sites (rate $\gamma$), empty with no neighbors (rate $\alpha$), empty with one neighbor (rate $\alpha \beta^{1}$), etc. Each category is assigned a weight based on the number of sites in the category and the relative rate of deposition/evaporation. The update process then procedes:

1. Randomly select one of the weighted categories.

2. Randomly select one site in the chosen category and flip its occupation state.

3. Reassign the updated site and nearest-neighbors to their new rate categories.

4. Advance the time: $t \rightarrow t+\Delta t$.

As with the previous algorithm, this process is repeated for a specified amount of time. Although the time will need to be rescaled to account for the exclusion of unsuccessful updates, this event-driven algorithm will reach the steady state much more efficiently regardless of the chosen parameters.

We use Monte Carlo simulations in the sections below to validate our model for two specific physical systems of different topologies: drug encapsulation using synthetic polymers with a tree-like structure, and  ionic self-assembled monolayers on a two-dimensional lattice. The simulations determine the final steady-state particle density and trace the system development as well. Simulations can also be used to investigate edge effects, different initial conditions, or a multitude of other aspects that are difficult to address analytically.

\section*{Applications}

We present two applications for this model utilizing separate geometries. The first is ionic self-assembly of nanoparticles for which we use a two-dimensional square lattice geometry. We then investigate drug encapsulation using dendrimers, which we model mathematically as Cayley trees.

\section*{2.4. Ionic self-assembly of nanoparticles and the two-dimensional lattice}

Self-assembly of nanoparticles is an important tool in nanotechnology and an active area of interdisciplinary research, with applications spanning a variety of fields such as optics, materials science, electronics, and nanomedicine \cite{decher}.  Conventional experimental techniques used to produce nanostructures fall under two categories: top down and bottom up. The top-down method starts with a bulk subtrate from which the material is progressively removed until the desired nanomaterial is obtained. Common top-down fabrication techniques are photolitography, electron beam lithography, and molecular beam epitaxy \cite{lindsay}. In the  bottom-up approach, nanostructures emerge from self-assembly of the system's components due to electrostatic or biochemical interactions \cite{decher}. 

One of the leading applications for nanoparticle self-assembly is bottom-up fabrication of thin films. Layer-by-layer self-assembly of nanoparticles on glass or polycarbonate substrates can be used to create thin-film structures known as Ionic Self-Assembled Monolayers (ISAM) \cite{iler}. Through repeated immersion in aqueous solutions of  appropriate ions, the self-assembly process deposits alternating layers of cations and anions on the substrate (Fig. 2). The resulting cation/anion bilayers, as shown on the right side of Fig. 2, form the basic building blocks for the film.

\begin{figure}[h] 
  \centering
  \includegraphics[width=4.5in,height=5.00in,keepaspectratio]{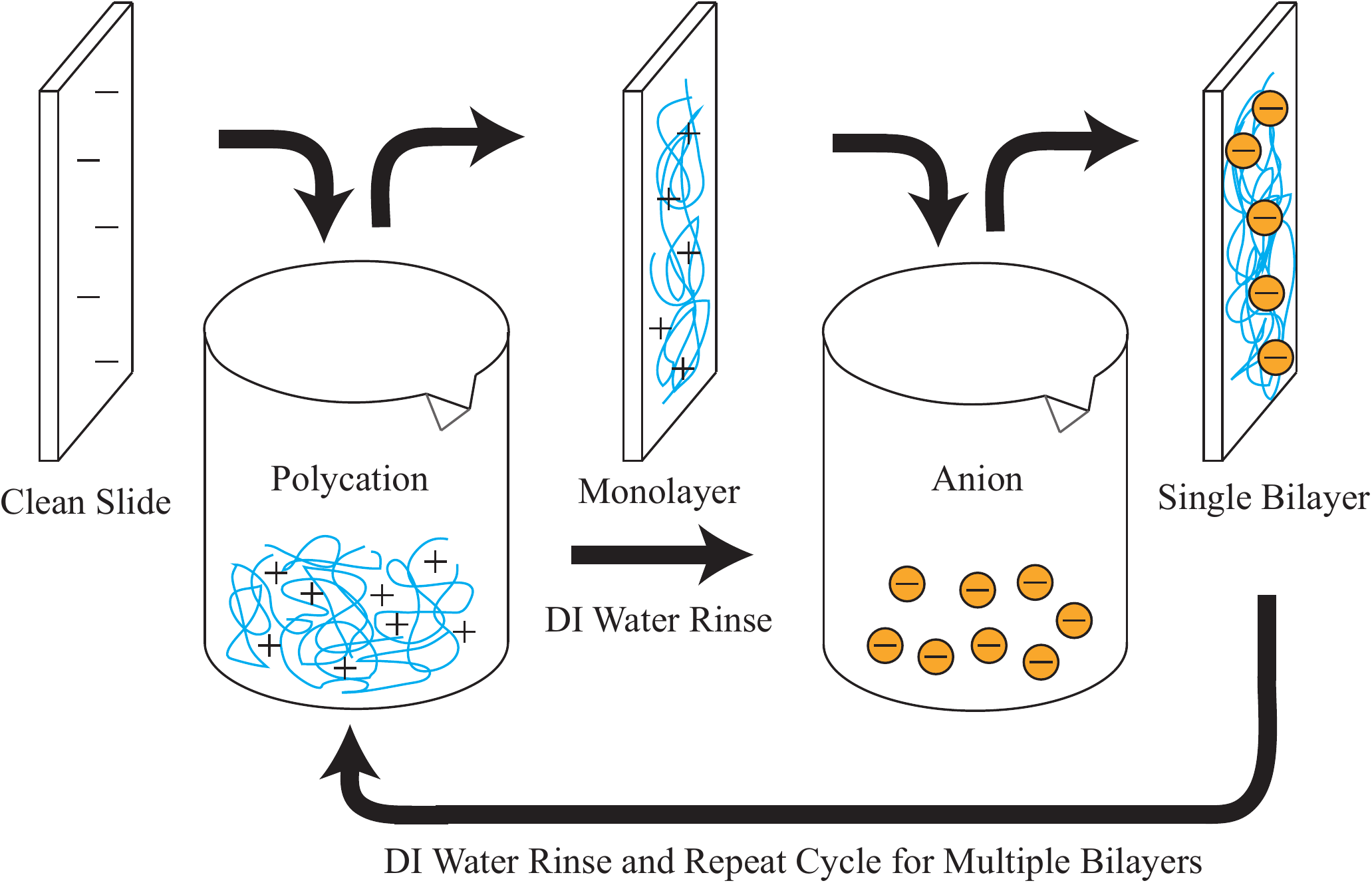}
  \caption{Illustration of the ionic self-assembly process.}
  \label{fig:Fig3}
\end{figure}

Because it is a dipping process, any exposed surface is homogeneously coated, allowing highly uniform, conformal coatings on irregular shapes.  The cation/anion bilayers may consist of either two polyelectrolytes (a polycation and a polyanion), a polyelectrolyte and a nanoparticle, or two different nanoparticles.  The thickness of a bilayer is a function of the diameter of the nanoparticle and the packing of the particles from layer to layer.  The optical properties of the resulting film can be tuned by the choice of nanoparticles and by the number of bilayers deposited.  A comprehensive review of the technique and its applications can be found in \cite{heflin}.  Although there are numerous studies on the subject of thin-film characterization \cite{films1,films2}, the goal of creating thin films with a graded index of refraction is still outstanding.  A study published by Yancey et al.\ \cite{ritter} shows that the coverage of the substrate plays an important role in tuning the index of refraction of the thin film.  The Maxwell-Garnett approximation \cite{optics book}, in fact, predicts that the index of refraction depends on surface coverage.

\section*{2.4.1. Ionic self-assembly: Experimental process}

In our experiments we deposited negatively charged spherical silica nanoparticles of nominal 40-50 nm diameter on negatively charged glass slides using poly(diallyldimethylammonium chloride) (PDDA) as polycation. The silica nanoparticles (SNOWTEX ST-20L from Nissan Chemical) were in a colloidal suspension at stable $pH=10.3$ and room temperature $T=21^{\circ}C$. The glass slides were cleaned under sonication, in three successive twenty-minute steps, with LABTONE detergent, 1N sodium hydroxide solution, and deionized water, and then dried with flowing nitrogen gas. The dipping time was ten minutes for each bilayer. We varied the concentration of the silica suspension by diluting it with deionized water. We examined the nanoparticle coverage of the substrate using SEM micrographs, in which deposited particles appear as light regions on a dark background. A sample SEM micrograph is shown in Fig. 3.  

\begin{figure}[h]
\centering
\includegraphics[width=4.5in,height=3.00in,keepaspectratio]{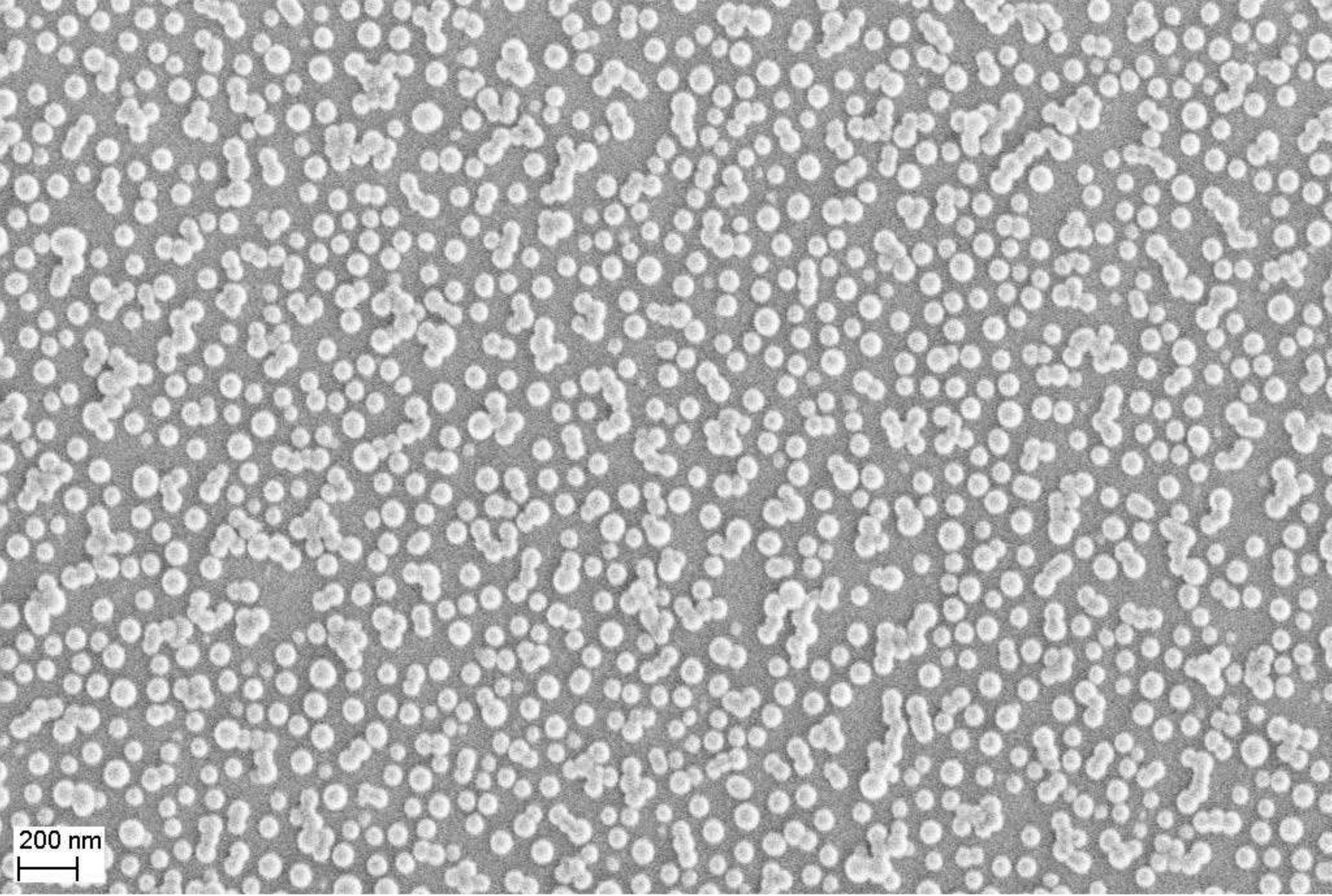}
\caption{A sample SEM micrograph at 25,000x magnification.  Nanoparticles deposited on the substrate appear as light regions.}
\end{figure}

We processed two single-bilayer micrographs for each concentration data point. Using an automated pixel-counting method we determined the average coverage of light pixels, representing presence of deposited particles. The experimental data is presented in section 2.4.4 alongside the analytical solution.

\section*{2.4.2. Ionic self-assembly: Theoretical modeling}

Due to the regularity of the cation layers in ISAM, we are able to model the nanoparticle deposition surface as a finite two-dimensional square lattice. Each site in the lattice represents a possible deposition site with a net positive charge.We consider the silica nanoparticles as charged monomers that attach to and detach from the lattice sites. Since the nanoparticles carry a net negative surface charge, electrostatic repulsion justifies direct application of our cooperative adsorption and evaporation model presented in Eq (1). Repulsion between nanoparticles causes the probability of deposition to decrease with each additional filled adjacent site, as addressed by a decreasing deposition rate in our model ($\beta < 1$). After relating the model parameters to physical ones, we are able to use our model to create films with specific particle densities. 

In Fig. 3, we see that the size of a particle is much less than the size of the slide. As such, edge effects due to the finite size of the slide should be negligible in the interior of the slide. Additionally, we see a uniform distribution of particles on the slide. These observations justify the assumptions of uniform coverage and translational invariance used in the mean field approximation.



\section*{2.4.3. ISAM: Computer simulations}
 
\begin{figure}[tbh]
\centering
\includegraphics[width=4.5in,height=3.00in,keepaspectratio]{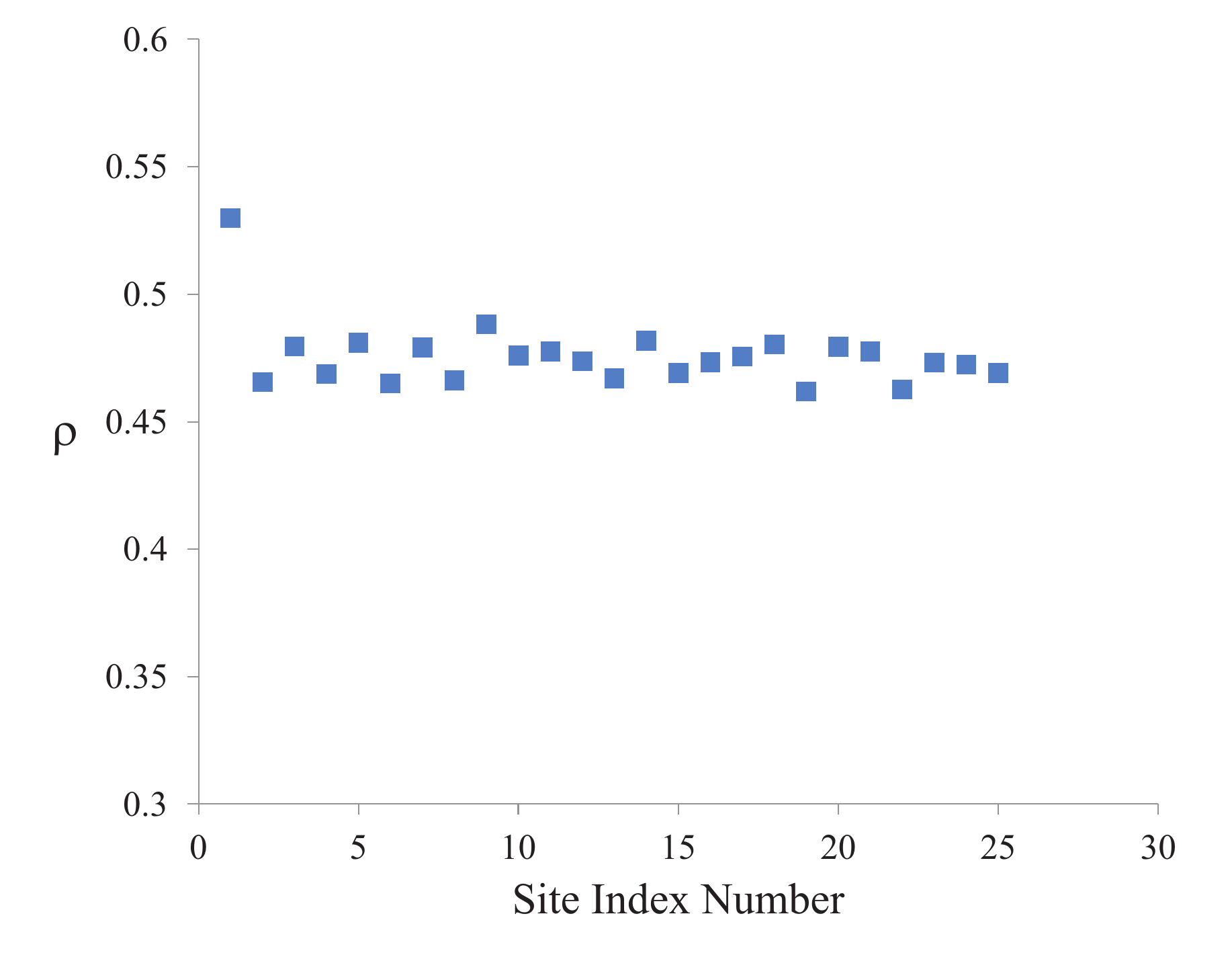}
\caption{Simulation results showing the the effects on the average site density near the left edge of the lattice. The sites are indexed with site 1 next to the edge.  The density shown is the column average of rows 31 to 90 on a $120\times120$ lattice with $\gamma=0.3$, $\alpha=1$, and $\beta=0.5$.\label{fig-edge}}
\end{figure}

We perform Monte Carlo simulations on a two-dimensional square lattice in order to investigate the dynamics of the CSAE process and evaluate the steady state solutions under various parameter regimes. Our simulations utilize a $120\times120$ two-dimensional grid onto which particles are both deposited at empty sites and evaporated from filled sites. In order to minimize edge effects, data is recorded for only the $100\times100$ matrix at the center of the larger lattice. The interior of the lattice is chosen instead of periodic boundary conditions; this choice mimics what is done experimentally. Only a small portion of the glass slide is analyzed, which is typically far away from the edges of the slide.  The edges in the simulations have a higher average site density due to the reduced number of neighbors.  Additionally, the average site density near the edges in the simulation decays rapidly to the average bulk density as shown in Fig.\ \ref{fig-edge} for the left edge.  Similar results are seen for the other edges.  Edge effects were not seen in the SEM micrographs, which is another reason we only consider the interior of the lattice.

Particles are deposited at empty cells with the rate $\alpha\beta^{\eta}$, where $\eta$ represents the sum of occupied neighboring sites. Particles evaporate from filled cells with the rate $\gamma$, which is independent of the state of neighboring sites. To update a site, a random site is chosen.  If the site is occupied, the particle will evaporate with rate $\gamma$.  If the site is empty, then it will become occupied with rate $\alpha\beta^\eta$.  We utilize an event-driven algorithm to make the simulation reach the steady state more efficiently. Starting with an empty lattice, we allow the system to reach steady state by waiting $1.44\times10^6$ site updates.  We then average the particle density at steady state over 100 realizations of the system.
 
For physical systems with repulsion between particles, the rate of deposition decreases when neighboring sites are occupied. We model this situation by choosing $\beta$ to be between zero and one for all simulations. A simple rescaling of time in Eq. (24) shows that the ratio of $\alpha$ to $\gamma$ controls the steady state density.  Therefore, we set $\alpha=1$ and vary $\gamma$ without loss of generality.

Fig. 5 presents the simulation data for steady-state particle density for our model on a 100$\times$100 square lattice for a variety of evaporation rates $\gamma$. This data shows excellent agreement with the density predictions from the mean field approximation in Eq. (25), reinforcing the validity of the approximation method.

\begin{figure}[h] 
  \centering
  \includegraphics[width=4.5in,height=7.00in,keepaspectratio]{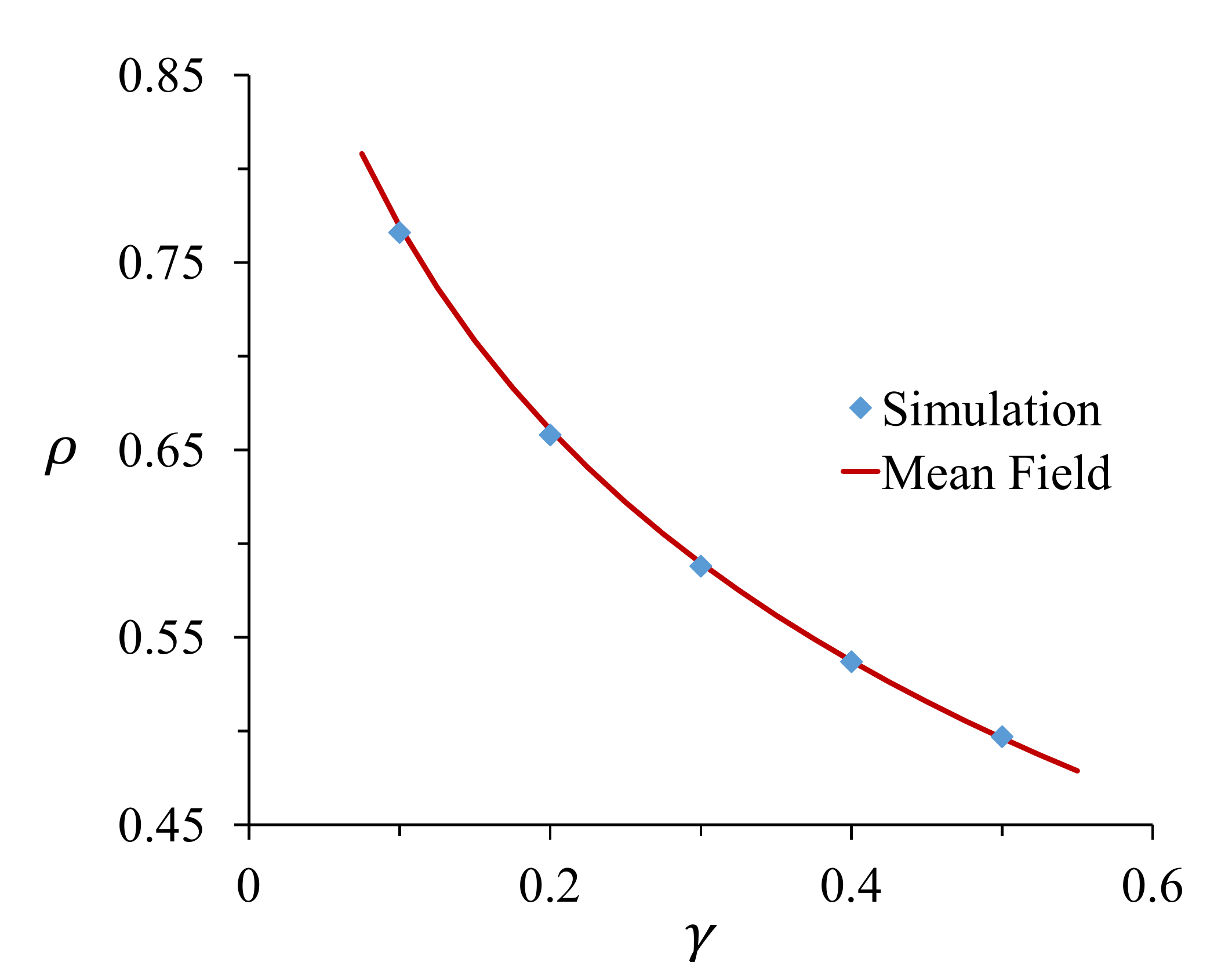}
  \caption{Sample Monte Carlo simulation results plotting density $\rho$ vs. evaporation rate $\gamma$ on a square lattice. Parameters used: $\alpha=1$, $\beta=0.7$.}
  \label{fig:Fig4}
\end{figure}


\begin{figure}[h]
\centering
\includegraphics[width=4.5in,height=3.00in,keepaspectratio]{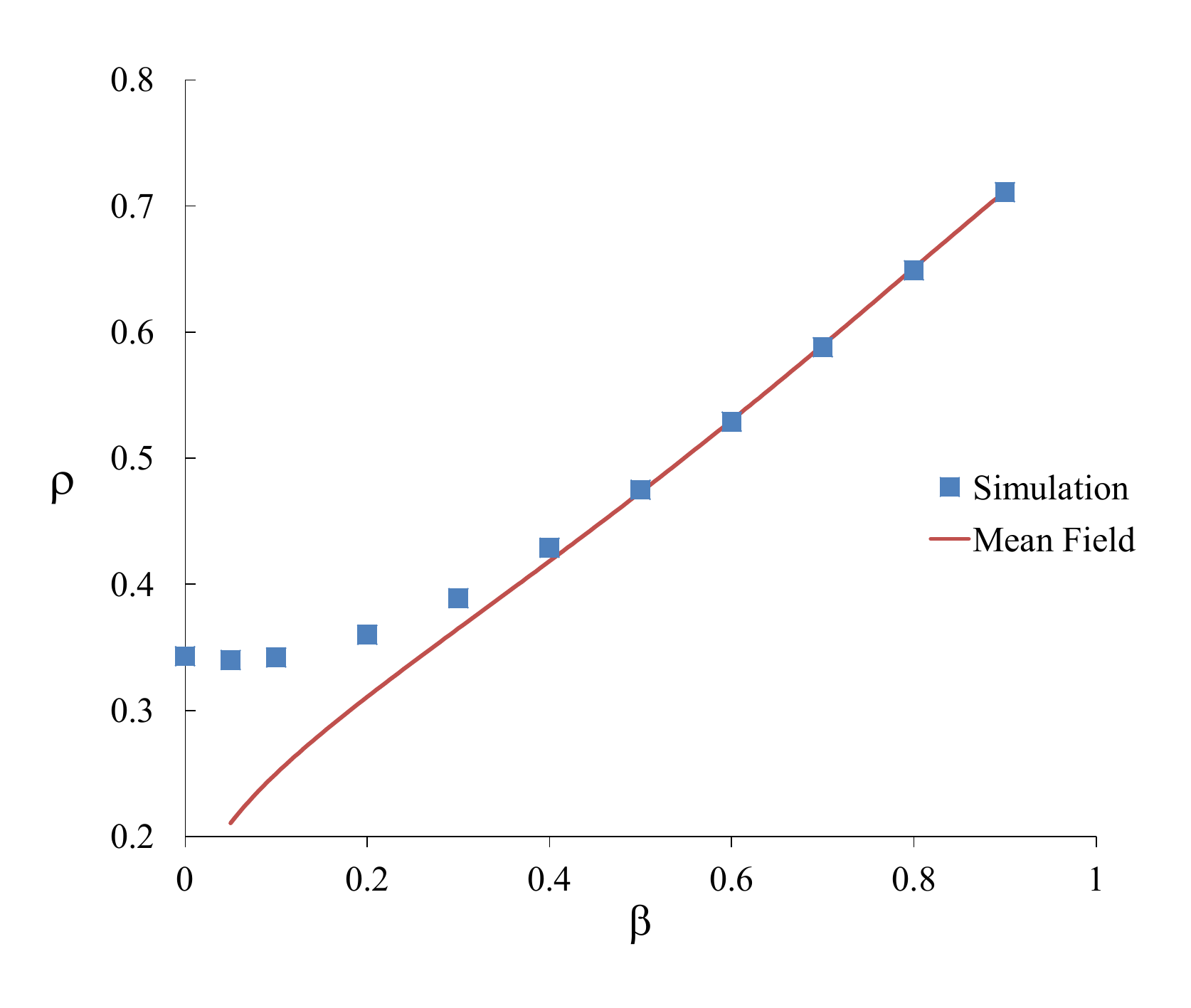}
\caption{Sample Monte Carlo simulation results plotting density $\rho$ vs. deposition interaction constant $\beta$ on a square lattice. Parameters used: $\alpha=1$, $\gamma=0.3$.}
\end{figure}

As seen in Fig. 6, the mean field result is in excellent agreement with the simulation results
and captures the relevant dynamics of this model for $\beta\ge0.4$. Additionally, we see that the density as a function of $\beta$ could be approximated by a linear function for $\beta\ge0.4$. At smaller values of $\beta$, the mean field theory fails to agree with the simulation due to stronger spatial correlation, which the theory neglects.  Since the simulation models an ISAM, we now look at connecting this linear dependence to a similar one found in the experimental data.

\section*{2.4.4. Ionic self-assembly: experiment and theory comparisons}\label{comparison}

Our CSAE model considers a simple case of deposition and evaporation of monomers and yields a transcendental equation for the particle density of the steady state that can be solved numerically. For the proposed model,  the equation (Eq. (25)) associated with the steady state is re-written here for convenience:

\begin{equation}\nonumber
\rho=\frac{\alpha\beta^{4\rho}}{\gamma+\alpha\beta^{4\rho}}.
\end{equation}

\begin{figure}[h]
\centering
\includegraphics[width=4.5in,height=3.00in,keepaspectratio]{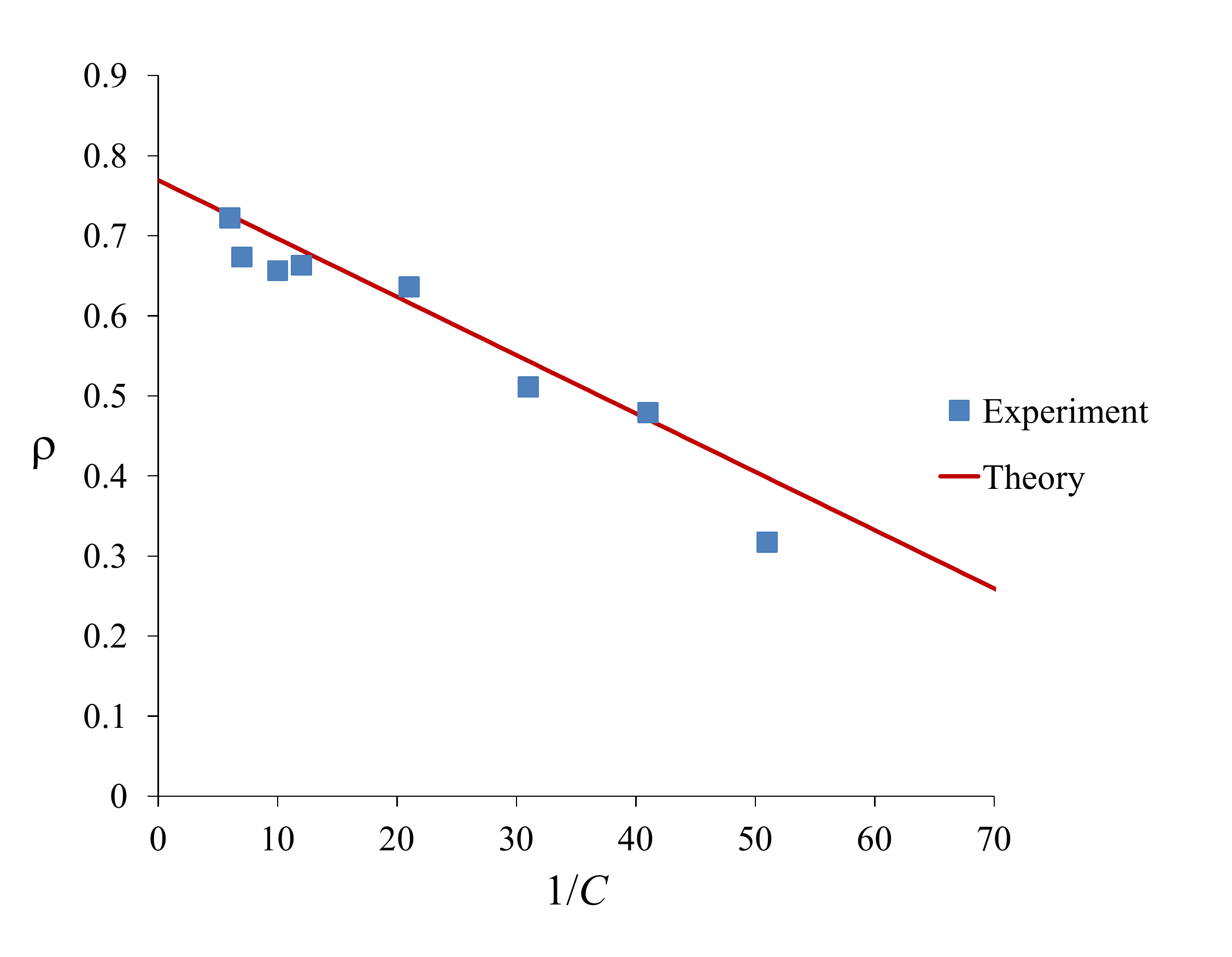}
\caption{Comparison of experimental data and theory for particle density as a function of the inverse of concentration of the colloidal suspension in arbitrary units. The equations associated with the linear fit are: i) theory (red line), $y=-0.0077 x+0.7655$; ii) experiment (blue squares), $y=-0.0078 x+0.7566$ with $R^2=0.94681$.  The theoretical fit is drawn from the numerical solutions to Eq. (25).\label{fig6}}
\end{figure}

The experimental data shows a linear dependence between the particle density and the inverse of concentration. Using our model, we found a relationship between the concentration of the nanoparticle suspension and the theoretical probability rate $\beta$. In particular, we found numerical solutions for the particle density $\rho$ in Eq. (25) for fixed $\alpha=1$ and $\gamma=0.3$, which match the experimental data shown in Fig.\ \ref{fig6}.

From this comparison, we conclude that, according to our model, for constant temperature,  the concentration of the nanoparticle solution  is a function of $\beta$:

\begin{equation}\label{concentration}
C=\frac{1}{75(1-\beta)}.
\end{equation}

for the chosen values of $\alpha=1$, and $\gamma=0.3$, or for any values of $\alpha$ and $\gamma$ for which the ratio $\frac{\gamma}{\alpha}=0.3.$

As long as this ratio is equal to $0.3$, the intercept of the theoretical and experimental lines is identical, and the difference between the two slopes for the experimental data fit and the theoretical fit is minimal.

For the model presented in this thesis, we also explored the case of the detachment rate being dependent on the number of neighbors, $\gamma^{\sum_{j \in NN}n_{j}}$. In the end, it seemed an unnecessary complication to consider such dependence, because the results showed that the model can be recast in terms of the ratio $\frac{\beta}{\gamma}$.
 
From an experimental point of view, this cooperative sequential adsorption model with evaporation can lead to interesting applications. The ability to predict or estimate the steady state coverage makes possible the prediction of the index of refraction \cite {optics book}, which is dependent on the overall particle density. A graded index of reflection is an outstanding goal in the creation of antireflective coatings. The model can also be modified for other lattice structures, such as Cayley trees with any coordination number $z$, with applications in modeling drug encapsulation in nanomedicine \cite{dendriHecht}.  The attachment and detachment rates can be chosen to reflect nearest-neighbor attraction ($\beta>1$) and repulsion ($\beta<1$).

 We conclude that we can directly relate our theoretical probability rate $\beta$ to the nanoparticle concentration using Eq. (30). 
 
 Several open questions may be addressed by extending our results.  The model presented matches well the particle density of the steady state found experimentally, but it doesn't capture the dynamics of the system on its way to the steady state. Experimental studies \cite{lvov} indicate that $90\%$ of the particle attachment happens in the first 30 seconds of the dipping process, followed by a slower approach to the final steady state. We plan to further study this time-dependent behavior both experimentally and theoretically. We will explore the dynamics of  our model when time-dependent attachment and detachment rates are being considered, in agreement with the experiment.  Our theoretical model can be generalized to include other aspects of the ISAM process, such as the presence of dimers and other particles of various shapes and sizes in the  colloidal suspension.

\section*{2.5. Cayley trees and drug encapsulation}

Nanomedicine is an emerging area of medical research that uses innovative nanotechnologies to improve the delivery of therapeutic and diagnostic agents with maximum clinical benefit while limiting harmful side effects. In recent years, self-assembly of nanoparticles has played an increasing role in nanomedical research in the context of drug delivery \cite{dendri1}. 

Dendrimers are highly branched polymers that consist of hydrocarbon chains with functional groups attached to a central core molecule.  Due to the precise control that can be exerted over their size, molecular architecture, and chemical properties, dendrimers have great potential in the pharmaceutical industry as effective carriers for drug molecules. These new synthetic polymers are able to carry both targeting molecules and drug molecules to cancerous tumors, minimizing the negative side effects of medications on healthy cells.

\begin{figure}[h] 
  \centering
  \includegraphics[width=4.5in,height=3.00in,keepaspectratio]{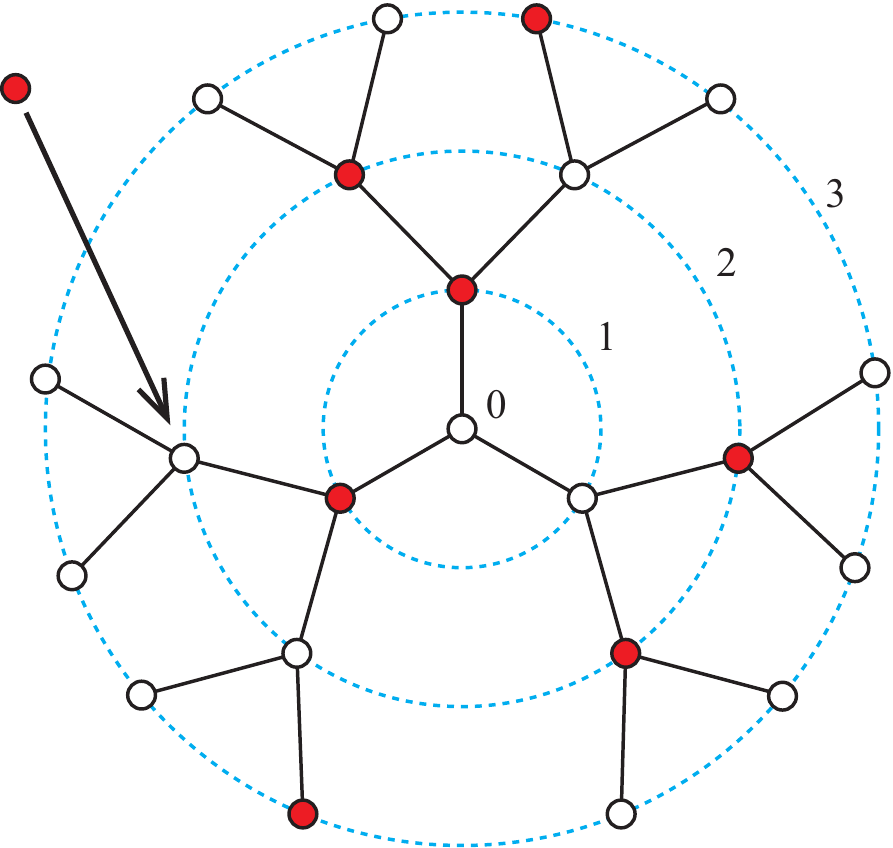}
  \caption{A sample configuration of a Cayley tree of coordination number $z=3$ with some nodes occupied. Generation numbers are labeled. The adsorption rate of a new particle depends on the number of occupied neighbors. For the considered node, the particle attempts to attach with rate $\alpha \beta$.}
  \label{fig:Fig1}
\end{figure}

There are two methods of drug delivery using dendrimers: \it encapsulation \rm of drugs within the cavities created by separate branches and \it attachment \rm of drug molecules to the outer functional groups. The potential load of each dendrimer carrier can be easily varied with an adjustment in branch multiplicity or dendrimer generation. The branches of the dendrimers  form a dendritic box around the encapsulated molecule, which can protect sensitive molecules from unfavorable physiological environments.  Dendrimer surface chains can form covalent or ionic bonds with anti-cancer molecules such as cisplatin \cite{dendri1, dendri2, dendri3, dendri4}. For both the encapsulation and the attachment process, the release of the drug molecules once they reach their target can be controlled with selective manipulation of the rate at which these dendrimer-drug bonds are degraded. It can happen instantaneously, or over the course of minutes or hours.

Analytical modeling of encapsulation and release of drug molecules is a very challenging task. The dendrimers are not rigid structures and they can change their shape and orientation depending on  a variety of factors, such as pH, temperature, and concentration of  drug molecule suspension. Over the years,  dendritic polymers and their interactions with drug molecules have been modeled using the following general methods: equilibrium and non-equilibrium molecular dynamics \cite{bosko, brodbeck}; equilibrium partition function calculations using the Ising model for localized electrostatic interactions between the drug molecules and the charged nodes of the dendrimer \cite{sun}; shell-like dendrimer models with a continuous and uniform charge distribution \cite{maiti}; cooperative sequential adsorption models solved using the empty interval method \cite{JSTAT}. 

 CSAE models are ideal for describing drug encapsulation and release because (i) the deposition process of the drug nanoparticles is stochastic and can be modeled by sequential adsorption models; (ii) the deposited drug nanoparticles are electrically charged, as are the substrate deposition sites, suggesting a cooperative model with deposition rates dependent on nearest-neighbor site occupation; (iii) the drug nanoparticles have a probability of detachment, which is incorporated in the model via an evaporation rate.

We model the dendrimer as a Cayley tree and address the drug encapsulation process using the CSAE model presented in Eq. (1). Mathematically, a Cayley tree of order $z$ is defined as follows \cite{ostilli}: Given a root vertex $0$, it is linked via $z$ edges to $z$ new vertices, forming generation  $\ell=1$. Each  $\ell=1$  vertex is linked to $z-1$ new vertices, forming generation $\ell=2$ and so on. Fig. 8 shows the first three generations  for a Cayley tree of coordination number $z=3$. Cayley trees are finite, with boundaries defined by the last generation. 

Our simple model considers the drug molecules to be  generic charged monomers that can attach themselves to the available oppositely charged nodes of the polymer. By controlling the model parameters, we can choose the density and spacing of the drug molecules that attach to the dendrimer.




\section*{2.5.1. Cayley trees: Particle density by generation}

In solving for the particle density on the Cayley tree geometry we employ the Ising model analysis introduced in section 2.1. While the standard mean field technique if often more convenient, the expanding nature of the Cayley tree makes the approximations used less appropriate. Each generation is larger than the previous one, often by a factor of three or four, so the number of sites on the edge of the tree makes up a significant portion of the total tree. This results in large edge effects and particle densities that vary by generation in the tree. We therefore investigate the particle density by generation on the Cayley tree and use those results to find the overall particle density. We will generally speak in terms of magnetization rather than particle density to match with the Ising model literature, but the magnetization $M$ is directly related to particle density $\rho$ by: $\rho = (1 + M) / 2$.

 Glauber presented the solution for the magnetization of a spin system in one dimension in \cite{glauber}. We generalize his method for a Cayley tree. We assume translational invariance within each generation of the Cayley tree: all spins within a specific generation are equivalent. We label the central node of the tree as $``n"$, and then each subsequent generation from $n$ to 1, with generation 1 being the outermost generation of the tree. We define the magnetization of generation $i$ as $q_{i}=<s_i>$. In terms of this magnetization, the particle density of generation $i$ is defined by: $\rho=(1+q_{i})/2$.
The time evolution of $q_{i}$ is derived \cite {Melin} to be:

\begin{equation}
\frac{dq_{i}}{dt}=-q_{i}(t)+B+<\tanh(K\sum_{j\in V(i)}s_{j})>+B<s_{i}\tanh(K\sum_{j\in V(i)}s_{j})>
\end{equation}
where $B=\tanh(h)$ reflects the effect of the external field.

This equation does not have exact solutions in higher dimensions, and in that situation, one has to use different approximations schemes. We present results for a Cayley tree of coordination number $z=4$  with arbitrary magnetic field $h$.

In order to be able to get a closed form for the system of equations, we use the series expansion approximation for $ \tanh(K\sum_{j\in V(i)}s_{j})
=C_1(\sum_{j\in V(i)}s_{j})+C_2(\sum_{j\in V(i)}s_{j})^3$, and find the coefficients $C_1=\frac{2}{3}\tanh(2K)-\frac{1}{12}\tanh(4K)$, and $C_2=\frac{1}{48}\tanh(4K)-\frac{1}{24}\tanh(2K)$. 

We also use the factorization approximation, $<\prod_{j \in NN}s_j> = \prod_{j \in NN} q_j$, which allows us to remove multi-spin correlations. With these approximations, the system of equations is:

\begin{eqnarray}
\frac{dq_{n}}{dt} &=&
q_{n}+4(C_1+10C_2)q_{n-1}+24C_2q_{n-1}^3
\nonumber
\\
&&+B(1+4(C_1+10C_2)q_nq_{n-1}+24C_2q_nq_{n-1}^3)
\nonumber
\\
\frac{dq_{i}}{dt} &=&
-q_{i}+(C_1+10C_2)(3q_{i-1}+q_{i+1})+6C_2(3q_{i-1}^2q_{i+1}+q_{i+1}^3)
\nonumber
\\
&&+B(1+(C_1+10C_2)(3q_iq_{i-1}+q_{i}q_{i+1})+6C_2(3q_iq_{i-1}^2q_{i+1}+q_iq_{i-1}^3))
\nonumber
\\
\frac{dq_{1}}{dt} &=&
-q_{1}+q_2\tanh(K)+B(1+\tanh(K)q_1q_2).
\end{eqnarray}

This system of equations can be solved numerically, and Figs. 9 and 10 present the associated particle densities per generation and for the entire tree for sample parameters. 

\begin{figure}[h]
\centering
\includegraphics[width=4.5in,height=3.00in,keepaspectratio]{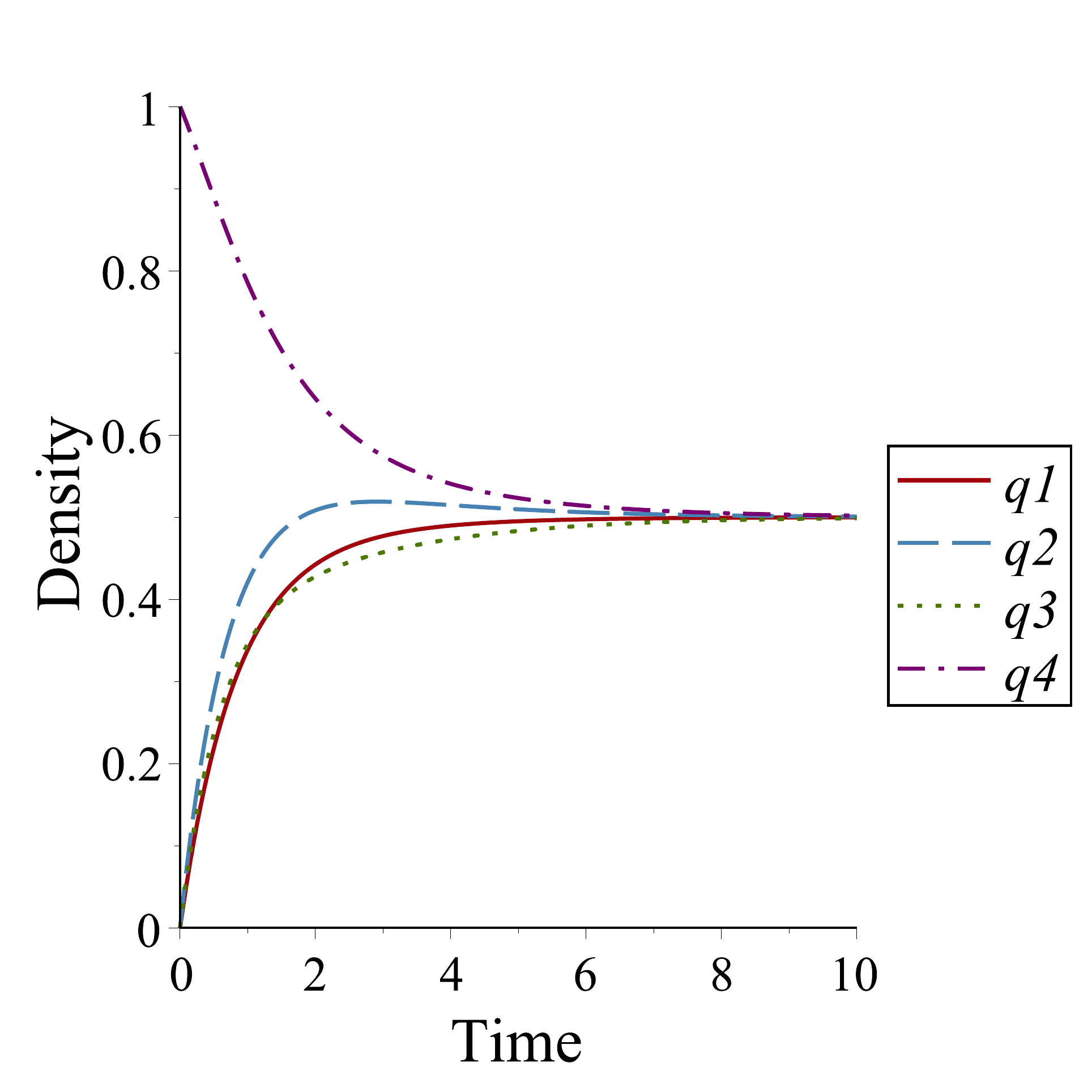}
\caption{\label{label}Density vs. time for each generation, with $ z = 4$, $\alpha = 1$, $\beta = 0.5$, $\gamma = 0.25$. Time is in arbitrary units.}
\end{figure}

\begin{figure}[h]
\centering
\includegraphics[width=4.5in,height=3.00in,keepaspectratio]{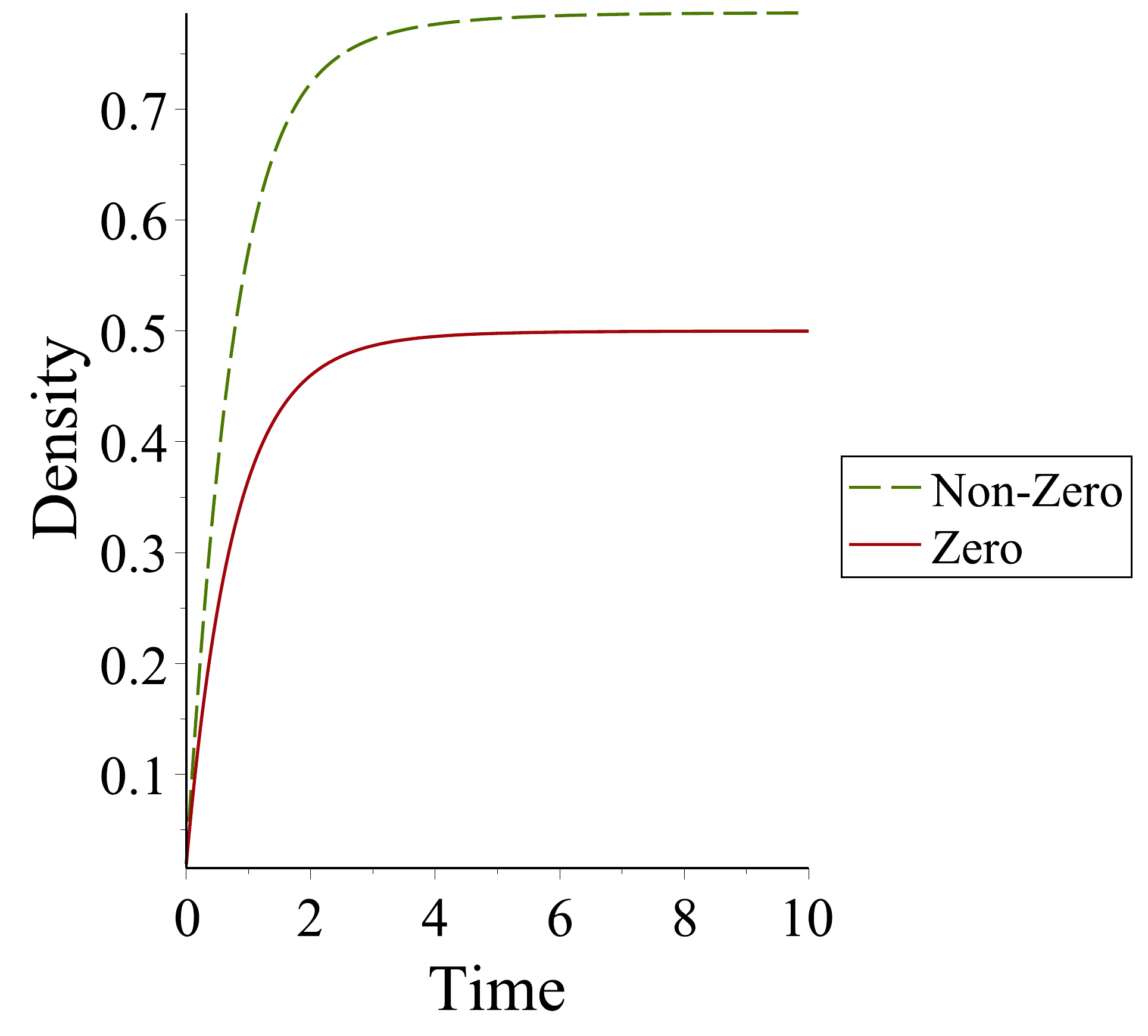}
\caption{\label{label}Density vs. time for entire tree for $z = 4$. Comparison for zero ($h=0$) and nonzero ($h=1$) external field. Time is in arbitrary units.}
\end{figure}

\begin{figure}[h]
\centering
\includegraphics[width=4.5in,height=3.00in,keepaspectratio]{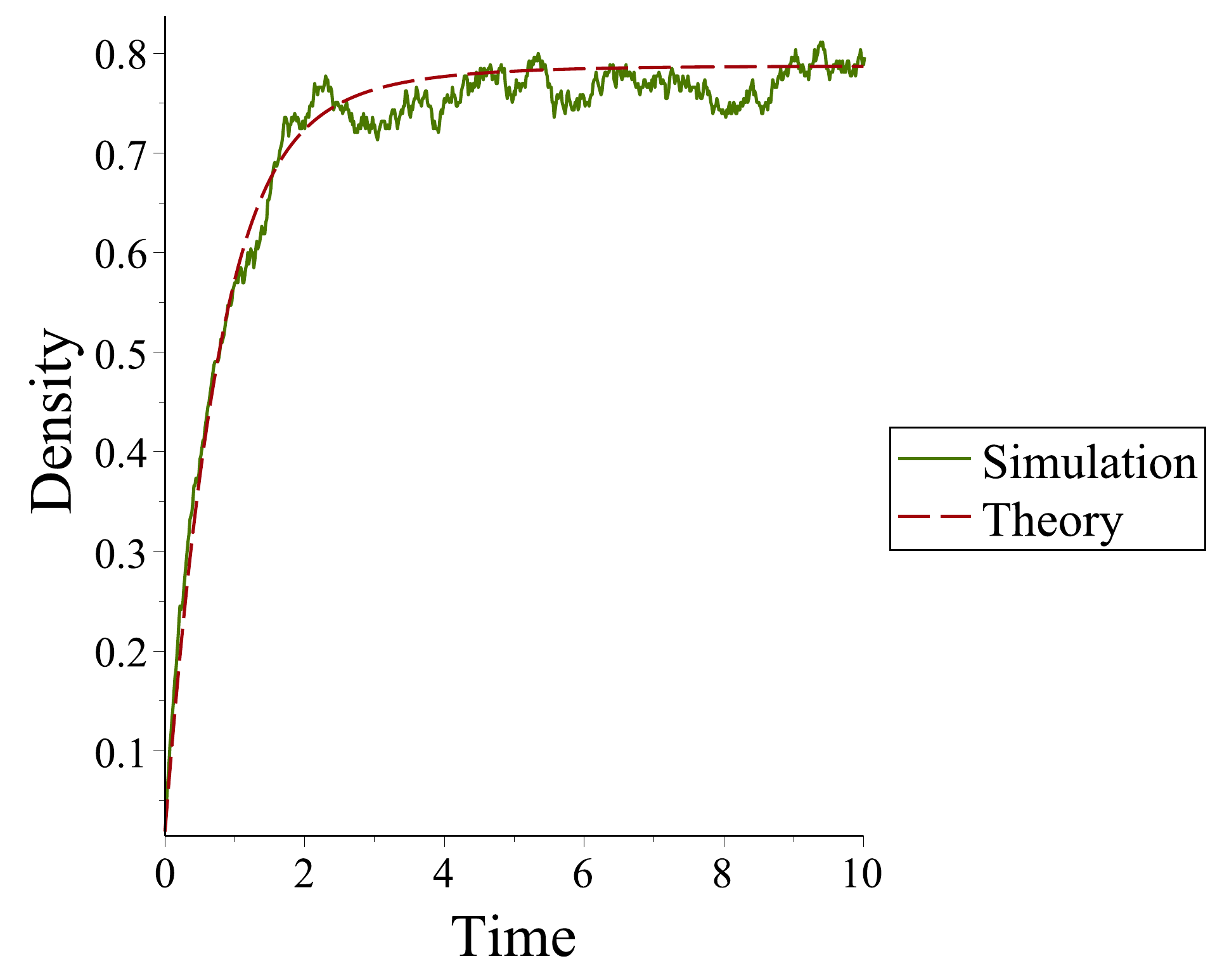}
\caption{\label{label}Comparison of theoretical solution (red, dashed line) and the simulation average over ten trials (green, solid line). Parameters used: $z = 4$, $\alpha = 1$, $\beta = e$, $\gamma = 1$. Time is in arbitrary units.}
\end{figure}

\begin{figure}[h]
\centering
\includegraphics[width=4.5in,height=3.00in,keepaspectratio]{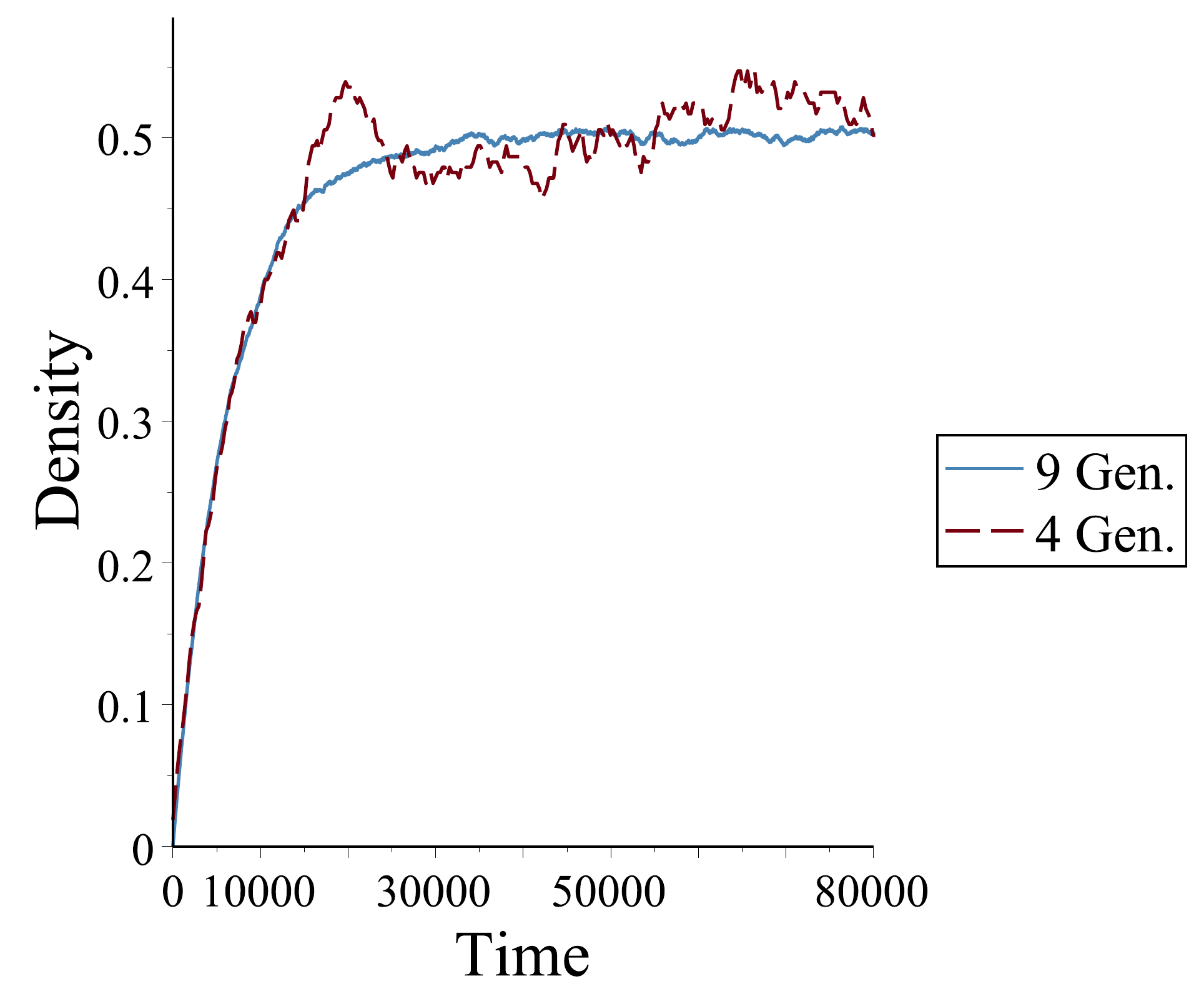}
\caption{\label{label}Comparison of simulations for 4-generation tree (red, dashed line) and 9-generation tree (blue, solid line) on an arbitrary time scale, for $z = 4$, $\alpha = 1$, $\beta = 1$, $\gamma = 1$.}
\end{figure}

We also matched our analytical results with Monte Carlo simulations on Cayley trees with coordination number $z = 4$ for a variety of parameter regimes, both with and without the presence of an external field. For an initially empty tree, particle density for the entire tree is the same as that predicted by our analytical solutions (Fig. 11). The good match between simulation results and analytical solutions suggests that simulations can be an effective tool for studying the dynamics of larger tree structures. A larger number of sites also decreases the impact of random variations on total particle density, making simulations more effective for larger systems than they are for small systems. We therefore investigate the time evolution of a 9-generation tree and compare it to our results for a 4-generation tree. As seen in Fig. 12, the density plots for the two trees are nearly identical when the arbitrary time for the larger tree is rescaled. This result suggests that our theoretical solutions will apply equally well to trees with more generations. The analytical model and computational simulations allow us to tailor the encapsulation process for specific types of dendrimers and encapsulated drugs. 

\section*{2.5.2. Cayley tree: Conclusions}

Theoretically, we found systems of differential equations describing the time-development of magnetization for each generation of Cayley trees with coordination number $z = 4$. We related these Ising model magnetization results to the particle density for our model. Computationally, we simulated the CSAE process on a Cayley tree and found excellent agreement between simulation results and theoretical predictions. This agreement validates our analytical solutions and supports the effectiveness of the simulations in mapping the dynamics of the system. Further studies could apply this model more directly to the drug encapsulation process or adapt it to address other systems such as self-assembled thin films, epidemic models, or social networks.

\section{Cooperative linear model}

In this section, we introduce our cooperative linear model and analyze it using many of the same techniques applied to the cooperative power model. While the cooperative power model discussed in the previous sections of the thesis has a wide variety of applications, a cooperative linear model is significantly easier to address analytically and can be used for many  of the same applications just as effectively.

As with the cooperative power model, we define the cooperative linear model on a uniform deposition surface where each site is connected to $z$ other sites, referred to as its nearest-neighbors. We consider singular nanoparticles which can both evaporate from the surface and attach to the surface with rates dependent on the number of occupied nearest-neighbors. The occupation state of site $i$ is determined by the occupation number $n_{i}$: $n_{i}=0$ represents an empty site and  $n_{i}=1$ represents an occupied site. 

We define the following transition rate for site occupation:

\begin{equation}
c(n_{i}\rightarrow(1-n_{i}))=  \gamma n_{i}
+  (1-n_{i})  \alpha \left(1- \beta \frac{1}{z} \sum_{j \in NN}n_{j} \right).
\label{eq: linear-transition}
\end{equation}

The first term in the transition rate accounts for evaporation: if a particle is present, it will evaporate with the probability $\gamma$. The second term addresses deposition: if a site is empty, a particle will attach with the rate $\alpha (1-\beta \eta / z$), where $\eta = \sum_{j \in NN}n_{j}$ is the number of occupied nearest-neighbors of site $i$. The parameters $\alpha$ and $\gamma$ set the relative rates for deposition and evaporation and $\beta$, restricted to the range $0 \le \beta \le 1$, controls the interaction with nearest-neighbors. For self-assembly with charged particles, electrostatic repulsion suggests that the presence of particles in neighboring sites will reduce the deposition rate. Higher values of $\beta$ increase repulsion effects, while the extreme case of $\beta=0$ models a situation with no interaction between neighboring particles.

With this transition rate, the number of particles on the lattice changes according to:

\begin{equation}
\frac{\partial n_i}{\partial t}=-\gamma n_i+(1-n_i)\alpha \left(1- \beta \frac{1}{z} \sum_{j \in NN}n_{j} \right).
\end{equation}

\section*{3.1. Linear model: Mapping onto the Ising model}

As with the power model, we can map the linear model onto the Ising model in order to use the known Ising model results. We follow the same process outlined in section 2.1 with the only difference being the subsitution of the linear model transition rate given by Eq. (\ref{eq: linear-transition}) instead of the power model transition rate given by Eq. (1). In the general case where each site has $z$ nearest neighbors, the coupling and external field constants become:

\begin{equation}
K\equiv\frac{J}{kT}=\frac{1}{2z} ln \left( \frac{1-\beta}{1-\beta/2} \right)
\end{equation}
\begin{equation}
h\equiv\frac{B}{kT} = \frac{1}{2} ln \left( \frac{ \alpha \left( 1 - \beta / 2 \right)}{\gamma} \right).
\end{equation}

These identifications again allow the use of established Ising model results for our model on different lattice types. It is interesting to note that the dependence on lattice shape denoted by $z$ is associated with the coupling constant for the linear model instead of with the external field constant as it was in the power model.

Since the Ising model results are cumbersome to work with analytically, especially after translating them into the particle deposition terms of our model, we will again turn to the mean field solution presented in the next section for the majority of our analysis.

\section*{3.2. Linear model: Mean field solution}

The mean field approximation can provide a method for solving for a rate equation for the particle density. We take the ensemble average of $\langle n_{i} \rangle$ and employ the mean field technique \cite{redner} to approximate the higher order correlations as $\langle n_in_j\rangle=\langle n_i\rangle\langle n_j\rangle$. We obtain:

\begin{equation}
\frac{\partial \langle n_i\rangle}{\partial t}=-\gamma \langle n_i\rangle
+  (1- \langle n_i \rangle )\alpha \left(1- \beta \frac{1}{z} \sum_{j \in NN} \langle n_{j} \rangle \right).
\end{equation}

Assuming translational invariance across the surface, the average site density $\rho_{i}=\langle n_{i} \rangle$ is the same as the total particle density $\rho = \sum_{i} p_{i}/N$, where N is the total number of sites. We can therefore write the rate equation in terms of the particle density $\rho$:

\begin{equation}\label{density-time-general}
\frac{\partial \rho}{\partial t}=-\gamma \rho+(1-\rho)\alpha (1 -\beta \rho).
\end{equation}

In the steady state $\partial \rho / \partial t = 0$, this equation is easily solved for the final particle density:

\begin{equation}\label{density-steady-state}
\rho= \frac{  \left( \alpha + \alpha \beta + \gamma \right) - \sqrt{ (\alpha + \alpha \beta + \gamma)^{2} -4 \alpha^{2} \beta}}{2 \alpha \beta}.
\end{equation}

\begin{figure}[h]
\centering
\includegraphics[width=4.5in,height=3.00in,keepaspectratio]{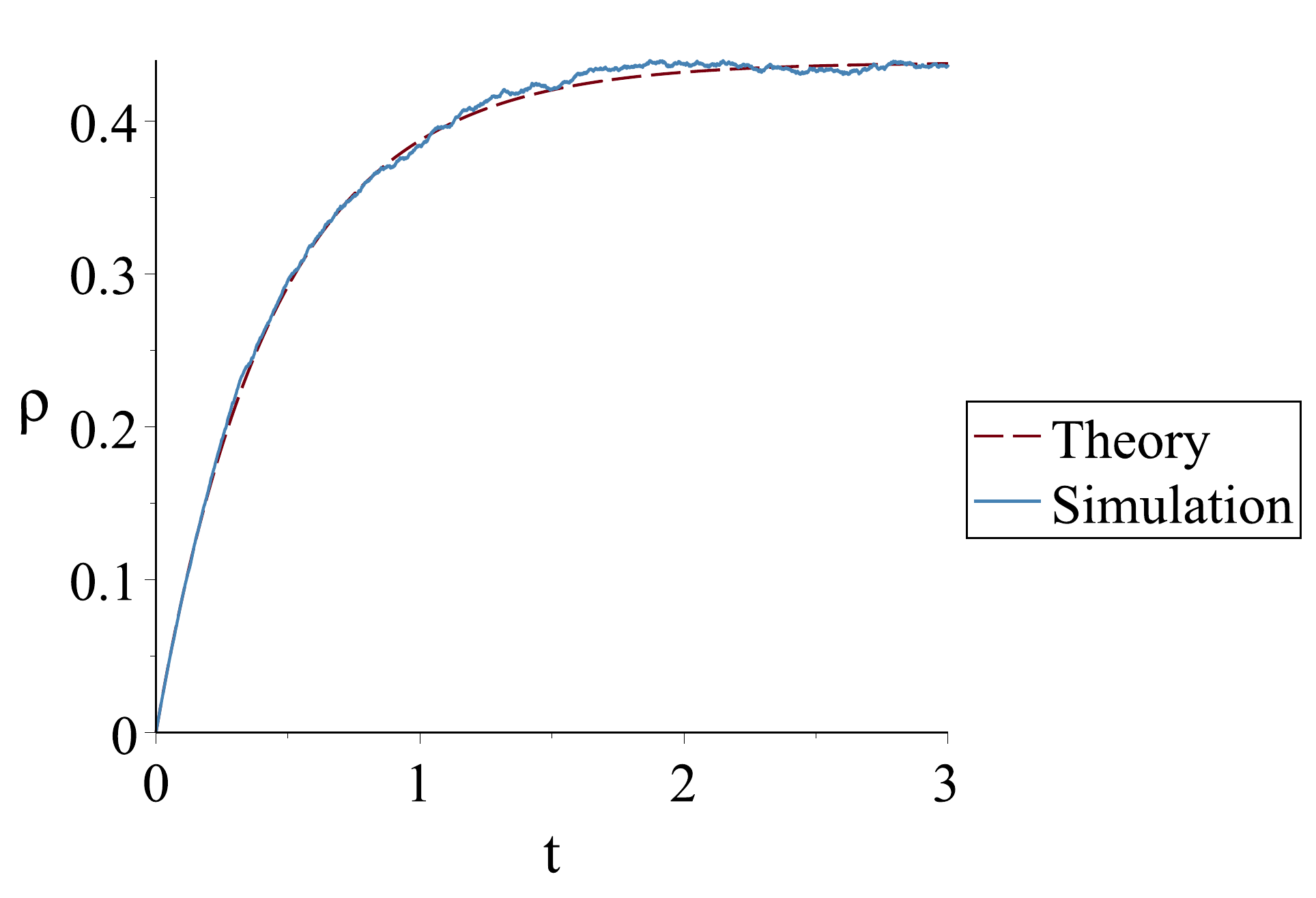}
\caption{Comparison of simulation (blue solid line) and theoretical (red dashed line) results for particle density over time in arbitrary units. Parameters used: $\alpha$ = 1, $\beta$ = 0.5, $\gamma$ = 1.}
\label{fig: Linear-time}
\end{figure}

Unlike the cooperative power model, this rate equation can also be solved exactly in the general case for the time-dependent particle density. The exact general solution is too cumbersome to be informative written here in full, but the solution is displayed graphically in comparison with simulation results in Fig. \ref{fig: Linear-time}.

It is interesting to note that the steady state densities given by Eq. (\ref{density-steady-state}) are nearly identical to those derived from the Ising model magnetization in the mean field approximation given in Eq. (28) with  $\rho=\frac{1+M}{2}$ and the coupling constants $K$ and $h$ given in Eqns. (35) and (36). 

\section*{3.3. Linear model: Simulations}

\begin{figure}[h]
\centering
\includegraphics[width=4.5in,height=3.00in,keepaspectratio]{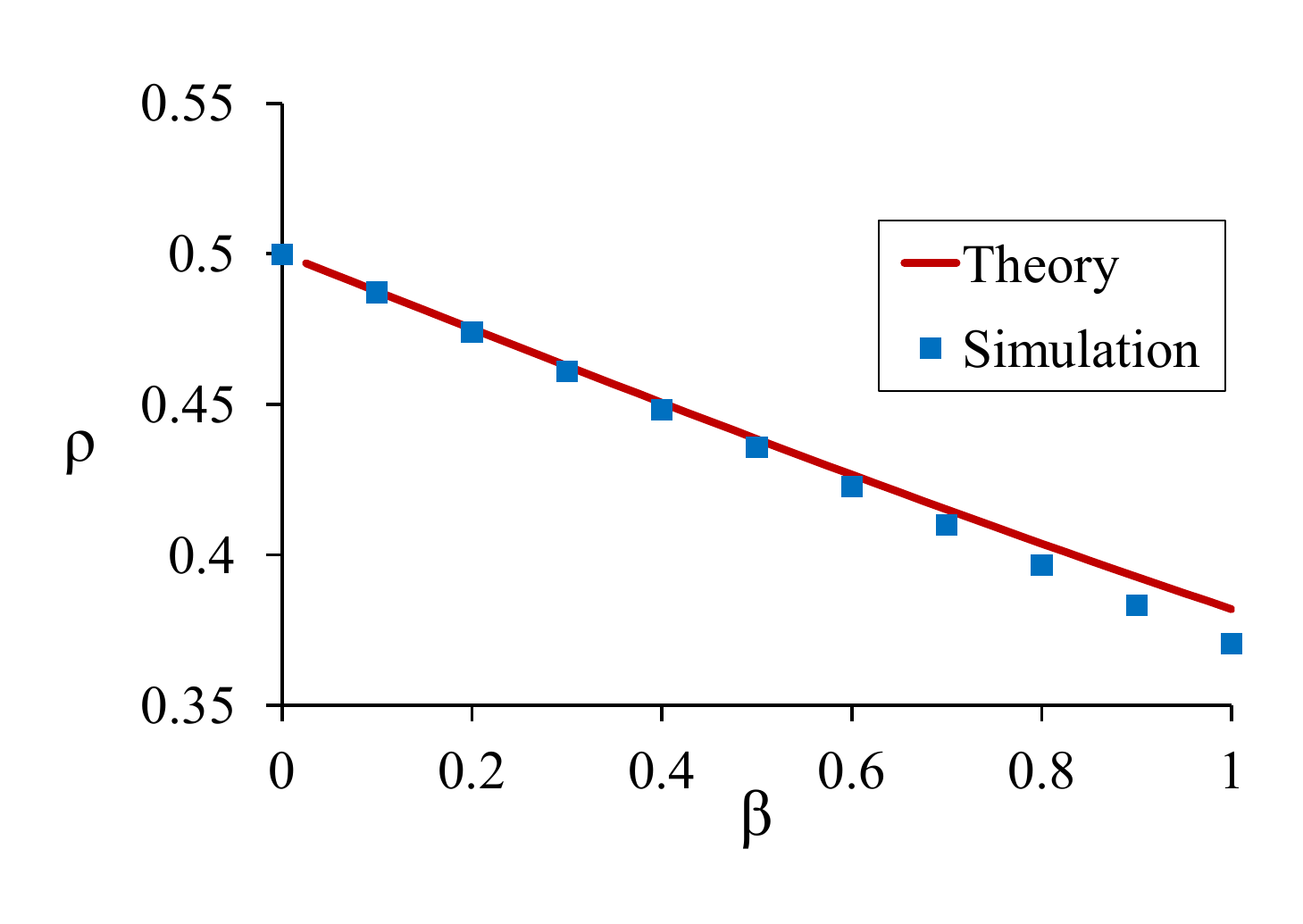}
\caption{Comparison of simulation (blue squares) and theoretical (red line) results for steady-state particle density as a function of $\beta$. Parameters used: $\alpha$ = 1, $\gamma$ = 1.}
\label{fig: Linear-beta}
\end{figure}

We utilize Monte Carlo simulations in order to further investigate the dynamics of the CSAE process. We model the deposition surface as a 120 $\times$ 120 square lattice. In order to minimize edge effects, we record data for only the 100 $\times$ 100 lattice section at the center of the larger lattice. 

The simulations proceed according to the transition rate given in Eq. (33) and utilize an event-driven algorithm in order to improve efficiency. Beginning from an empty lattice, we allow the simulation to proceed for 1.44 $\times$ $10^{6}$ site updates in order to ensure that a steady state is reached. We average the particle density results over 100 realizations of the system.

We found excellent agreement between simulations and theoretical results for both the time development of the system and steady-state particle densities as shown in Figs. \ref{fig: Linear-time} and \ref{fig: Linear-beta}. 

\section*{3.4. Linear model: Comparison with experiment}

We can analyze a potential use for this model by applying it to the experimental results presented in section 2.4. Comparing our theoretical model with the experimental data, we find the nanoparticle concentration $C$ can be related to a function of the interaction parameter $\beta$:

\begin{equation}
C = \frac{1}{23 \beta + 5}.
\label{eq: linear-exp}
\end{equation}

The parameters $\alpha$ = 1 and $\gamma$ = 0.4 were chosen to match experimental data for the concentration relation. A graphical comparison of our experimental and theoretical results is presented in Fig. 15.

\begin{figure}[h]
\centering
\includegraphics[width=4.5in, height=3.00in, keepaspectratio]{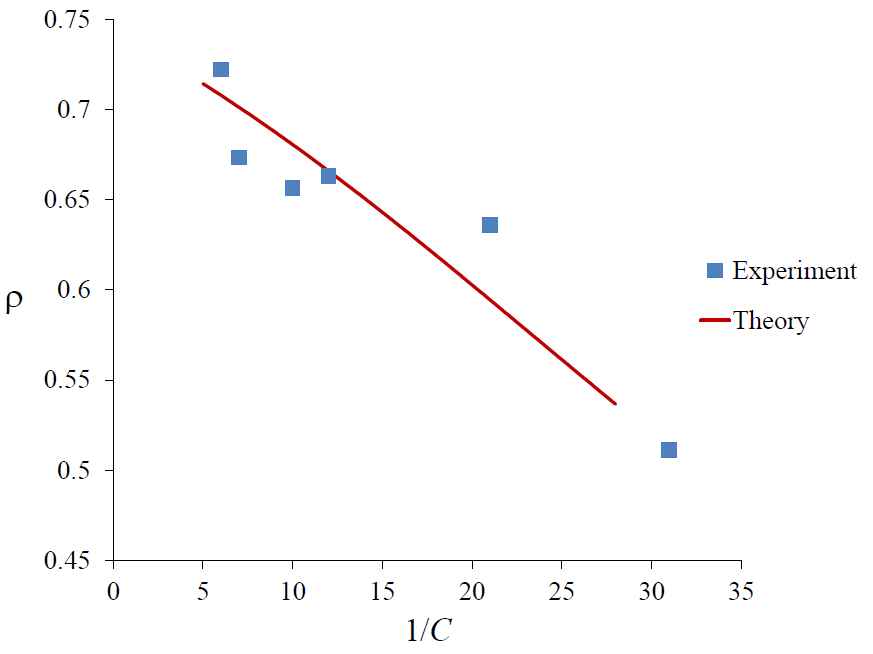}
\caption{Comparison of experimental data and theoretical results for steady-state particle density as a function of the inverse of nanoparticle concentration in arbitrary units. The linear fit equations are: i) theory (red line), $y=-0.0078x+0.7582$ with $R^{2}=0.9988$; ii) experiment (blue squares), $y=-0.0078x+0.7566$ with $R^{2}=0.9468$. Parameters used: $\alpha$ = 1, $\gamma$ = 0.4.}
\label{fig: Linear-exp}
\end{figure}

As seen in the Fig. 15, the analytical results for the cooperative linear model can provide an excellent fit for the ISAM experimental data for varied nanoparticle concentration. As with the cooperative power model, the parameter $\beta$, representing the strength of particle interactions, is directly related to the concentration of nanoparticles. The theoretical predictions, however, are restricted to the density range shown by the theoretical curve in Fig. 15 for the chosen values of $\alpha$ and $\gamma$. 

\section*{3.4. Linear Model: Comparisons and conclusions}

The similar results for the cooperative linear model and the cooperative power model merit a comparison. The linear model presented in this section has the advantage of being much simpler to work with analytically. It has direct solutions for both the steady-state and time-dependent particle densities where the cooperative power model requires numerical approximations. The downside of the cooperative linear model is that the solution presented above (Eq. (\ref{eq: linear-exp})) is restricted in particle density range. The constants chosen to match the experimental data can be varied to match the data for different particle density ranges, but we do not have a single solution that applies to the entire range of experimental data for the cooperative linear model. This issue is not encountered with the cooperative power model, where the single result (Eq. (\ref{concentration})) applies to all of our experimental data.

Additional investigation into the link between the interaction parameter $\beta$ and nanoparticle concentration $C$ could lead to a more thorough understanding of the relationship and a provide general solution for the cooperative linear model that applies over a larger particle density range. Future studies could also investigate the applications of this model to different geometries or physical systems. 

\section{Total lattice cooperative model}

This section introduces another stochastic model with properties and analytical techniques significantly different from the previous two models. Instead of cooperative effects limited only to nearest-neighbors, the attachment rate depends on the overall number of  particles already present in the system. In effect, every site in the system acts as the neighbor for every other site. The model is quite general, is applicable to all dimensions and topologies, and can describe a variety of two-state physical systems. We use both mean field theory and matrix theory to find solutions for the particle density and probabilities of having a set number of particles present in the system. We compare our analytical results to  Monte Carlo simulations and to experimental data on the  self-assembly of charged nanoparticles on glass substrates \cite{iler}.

\section*{4.1. Total lattice model: Model description}

Our total lattice cooperative model is defined on a general lattice  of arbitrary topology  (rectangular grids, Cayley trees, etc.) of $N$ sites.  Each site of the grid has two states: empty or filled. Empty sites are filled at a rate $\alpha_i$; filled sites are emptied at a rate $\beta_i$.  These rates are functions of the total number $i$ of filled sites in the lattice. In order to mimic the cooperative effects due to electrostatic repulsion during the ionic self-assembly of nanoparticles, we consider the attachment rates to decrease as the number $i$ of  filled sites increases, and the detachment rates to increase with $i$. The functions picked for these rates can be modified easily depending on the physical situation considered. We consider here a linear case:
\begin{eqnarray}
\alpha_i&=&\alpha(N-i) \\ \nonumber
\beta_i&=&\beta i
\end{eqnarray}
with $\alpha$ and $\beta$ positive constants. This model has the virtue of simplicity and can be solved using the mean field approximation, but exhibits sufficient complexity to be useful as a standard of comparison for experimental results as well as analytic and computational models that include more complex rate assumptions.  

As with the other models presented in this thesis, we can find a rate equation for the density of filled sites  $\bar{\rho}$ most easily using the mean field approximation:

\begin{equation}
\frac{d \bar{\rho}}{dt} = \alpha (1-\bar{\rho}) ^2-\beta \bar{\rho} ^2,
\end{equation}
with general solution
\begin{equation}
\bar{\rho}(t)={\frac {\alpha-\tanh \left( t\sqrt {\alpha\,\beta}+A\sqrt {\alpha\,
\beta} \right) \sqrt {\alpha\,\beta}}{\alpha-\beta}}
\end{equation}
for positive $\alpha$, $\beta$ and $\alpha \neq \beta$. When the two rates are equal, $\bar{\rho }\left( t \right) =1/2+A\,{{\rm e}^{-2\,\alpha\,t}}$, and the surface coverage settles at $50\%$.

If cooperative effects are not being considered (constant attachment and detachment rates), the solution for the particle density is:
\begin{equation}
 \bar{\rho} = A \exp\left( -(\alpha+\beta) t \right) + \frac{\alpha}{\alpha+\beta}.
\end{equation}
In each case, the coefficient $A$ is determined by the mean density at $t=0$.

Moving beyond the mean field result, we let $Q_i$ represent the time-dependent ensemble-average probability that exactly $i$ sites of the lattice are filled. This obeys the master equation:
\begin{equation}
\frac{d}{dt}\, Q_i = -((N-i) ^p\alpha + i^{p}\beta) Q_i + (N-i+1)^p \alpha Q_{i-1} + (i+1) ^p\beta Q_{i+1}  \label{rateeq}
\end{equation}
where $p=1$ for constant attachment and detachment rates, and $p=2$ for variable rates with linear dependence on the number of filled sites at time $t$. This equation can be solved exactly using matrix theory or the generating function method for $p=1$. 

The general time-dependent solution is given by $Q_i = \sum_{k=0}^N c_k E_{ik} \exp(\lambda_k t) $, where $\lambda_k$ is the $k$-th eigenvalue of the associated matrix, and $E_{ik}$ is the $i$-th component of its $k$-th eigenvector.  The derivation of these eigenvalues and eigenvectors for the $p=1$ case has been presented in a recent article \cite{fonseca}.

\begin{figure}[h]
\centering
\includegraphics[width=4.5in,height=3.00in,keepaspectratio]{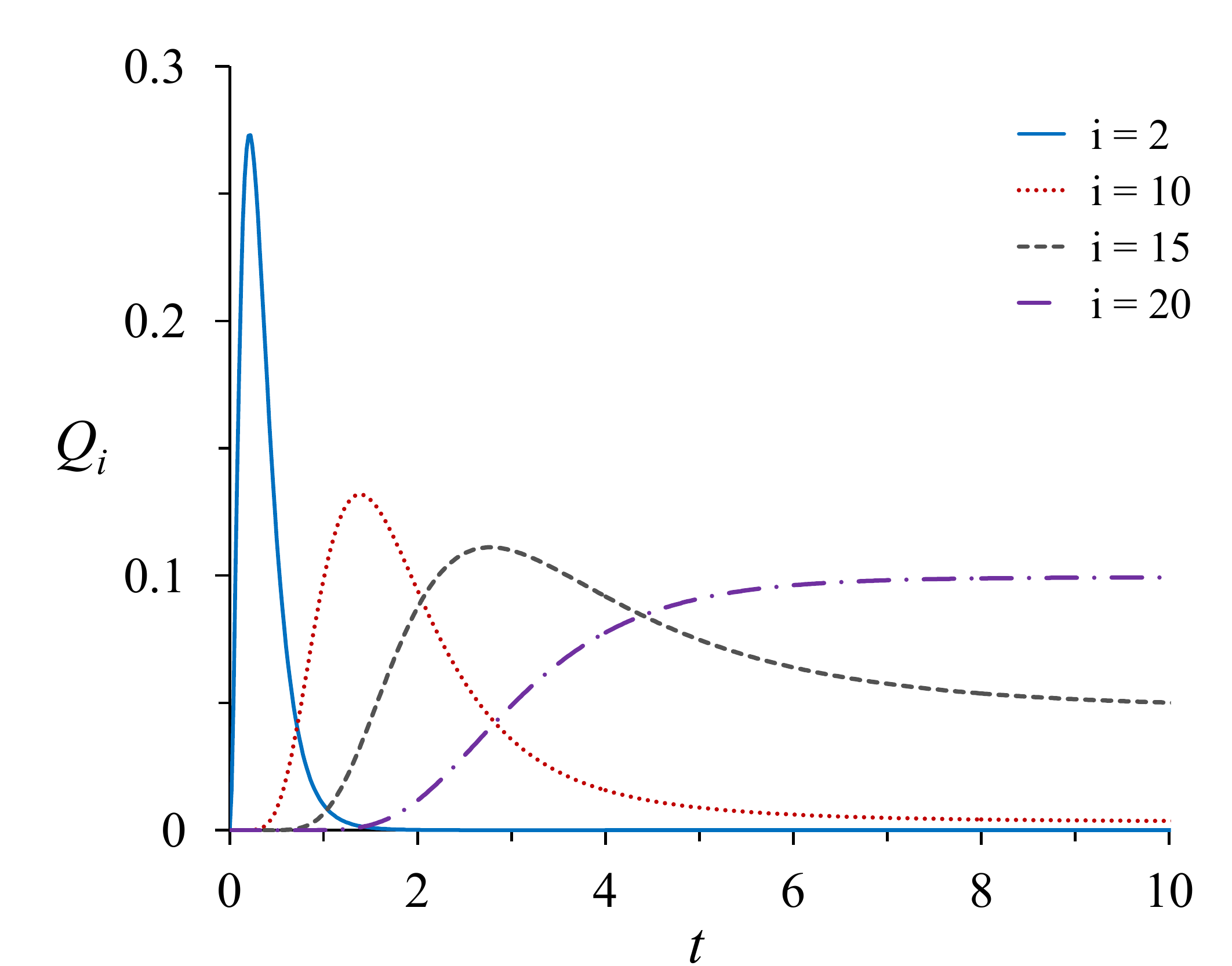}
\caption{\label{label}The probability $Q_{i}$, as a function of time, that exactly $i$ sites are filled in a case of $N=100$ initially filled sites and $\eta = 4$.}
\end{figure}

We show in Fig. 16  the probability, as a function of time, that exactly $i$ sites are filled for several values of $i$, in a case with $N=100$ initially empty sites, using $\frac{\beta}{\alpha} = 4$, for constant attachment and detachment rates.  As time progresses, each $Q_i$ for $0 < i < 20$ at some point becomes the dominant term, spikes, and then diminishes. The expected steady state  at $i=20$ does not spike, but rather levels off as $t$ increases. $Q_{20}$ does not become unity because the deposition and evaporation will continue, meaning that other states will remain possible at all times, with states nearest to $i=20$ having higher probabilities than others.

\section*{4.2. Total lattice model: Computer simulations and theory}

We have created Monte Carlo simulations to further investigate the stochastic particle adsorption process. We present results from simulation runs on a $100\times100$ two-dimensional grid using two sets of attachment and detachment rates. The first series of simulations uses the variable attachment and detachment rates presented in Eq. (41). The second series, presented for comparison, uses constant attachment and detachment rates. 
\begin{figure}[b]
\centering
\includegraphics[width=4.5in,height=3.00in,keepaspectratio]{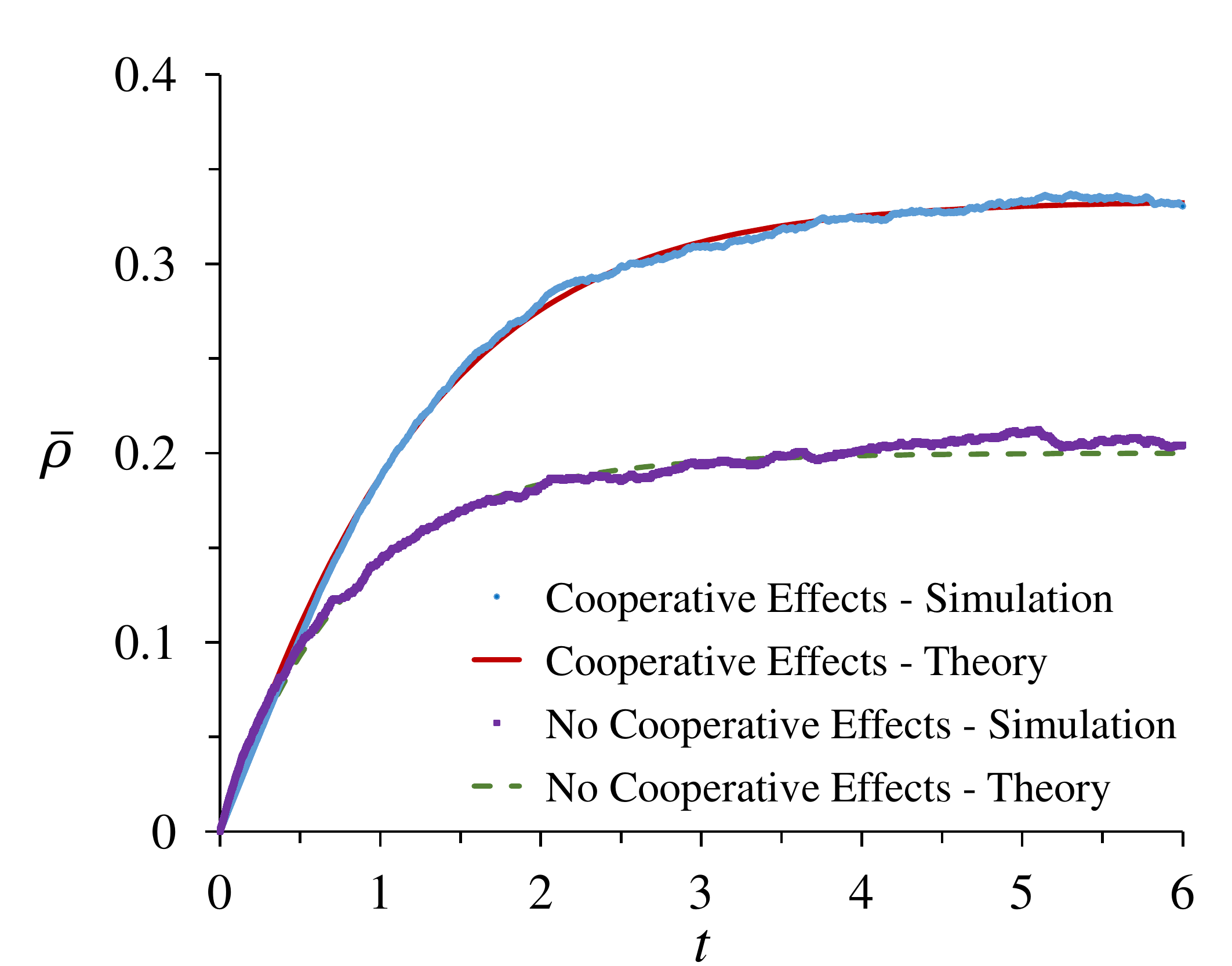}
\caption{Mean particle density as a function of time for $\eta=4$, with and without cooperative effects. Time is in arbitrary units.}
\label{fig:RSAE Density-Time}
\end{figure}

\begin{figure}[b]
\centering
\includegraphics[width=4.5in,height=3.00in,keepaspectratio]{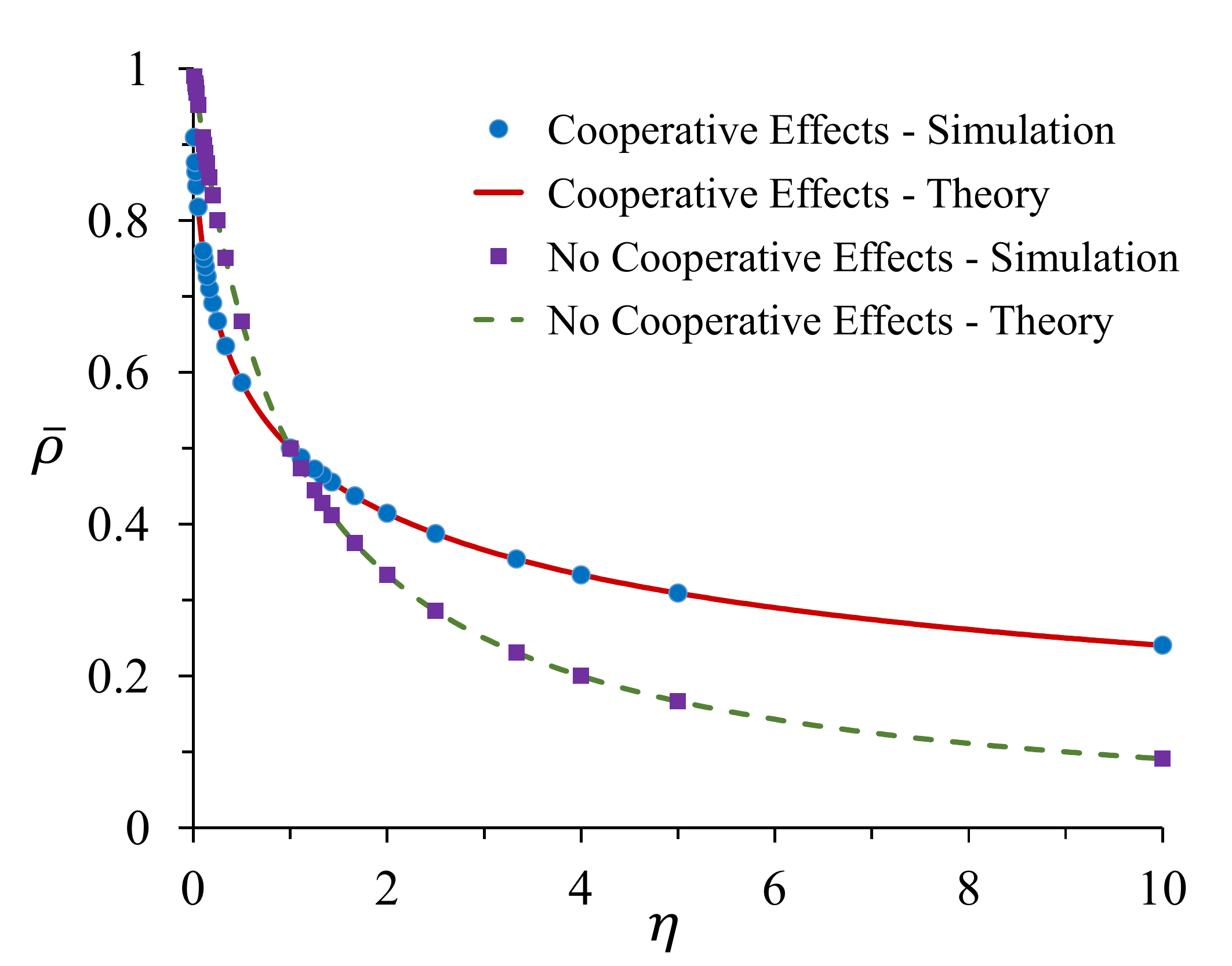}
\caption{Mean particle density for the steady state as a function of $\eta$, with and without cooperative effects.}
\label{fig:RSAE Density-SS}
\end{figure}

We utilize an event driven algorithm for both series of simulations. We allow the simulation to proceed for $1.44\times10^{6}$ site updates in order to ensure a steady state is reached. We record the particle density over time (Fig. 17) as well as the steady state particle density averaged over 100 realizations of the system (Fig. 18). These plots show the excellent agreement between the analytical solutions and the computer simulations and validate our use of the mean field approximation in deriving the analytical results.

\section*{4.3. Total lattice model: Experiment and theory}

As with the previous two models, we investigate a potential application of the total lattice model to the experimental ionic self-assembly data presented in section 2.4. Comparing the theoretical model to the experimental results, we find that the nanoparticle concentration $C$ can again be related to the deposition rate ($\alpha$ in this model):

\begin{equation}
C = \frac{\alpha}{45} + 0.009.
\end{equation}

The detachment parameter $\beta$ = 1 is chosen for the concentration relation to match experimental data. A graphical comparison of the theoretical predictions and experimental data is presented below in Fig. \ref{fig: Total-exp}.

\begin{figure}[h]
\centering
\includegraphics[width=4.5in, height=3.00in, keepaspectratio]{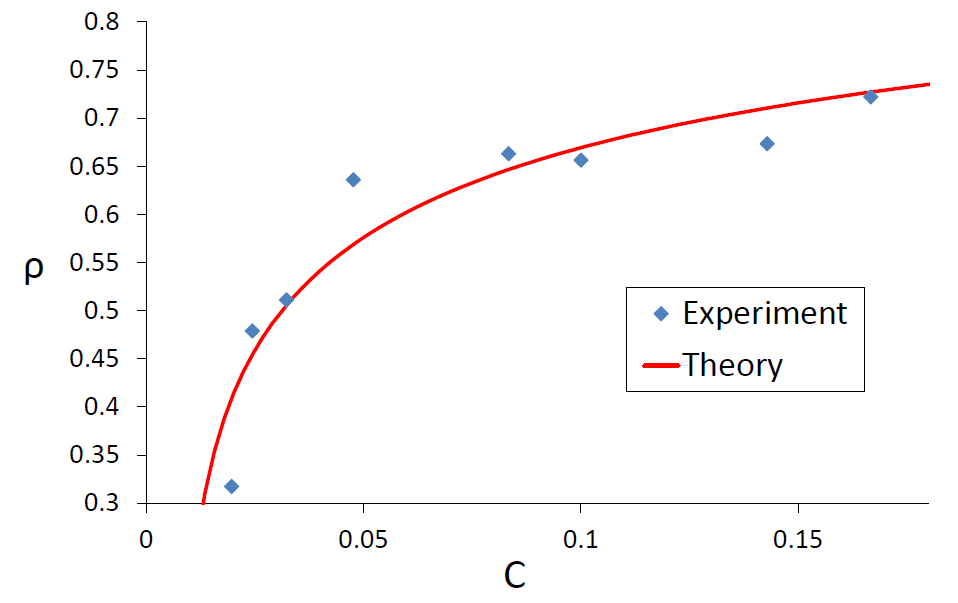}
\caption{Comparison of experimental data and theoretical results for steady-state particle density as a function of the nanoparticle concentration in arbitrary units. The fit equations are: i) theory (red line), $y=0.151 ln(x) + 1.0107$ with $R^{2}=0.9747$; ii) experiment (blue diamonds), $y=0.1517 ln(x) + 1.0118$ with $R^{2}=0.834$. Parameter used: $\beta$ = 1.}
\label{fig: Total-exp}
\end{figure}

As shown in Fig. \ref{fig: Total-exp}, the total lattice cooperative model can be used to model ionic self-assembly of silica nanoparticles. The total lattice model, however, does not fit the experimental data as well as our previous two models. We observe significantly lower $R^{2}$ values and increased variation between the theoretical and experimental curves. This increased deviation is expected due to the localized nature of the electrostatic repulsion between silica nanoparticles. The total lattice cooperative model approximates well the ionic self-assembly process for the case of low particle density or low levels of interaction, but it cannot model strong localized forces. 

\section*{4.4. Total lattice model: Conclusions}

The total lattice cooperative model has significant advantages and disadvantages when compared to the cooperative power model or the cooperative linear model. Its main advantage is the ease of working with it analytically. We were able to calculate and present exact analytical results for the time-dependent probability that $i$ sites of a $d$-dimensional lattice are filled for the case of constant attachment and detachment rates. We also calculated the mean coverage of the lattice for both constant and variable attachment and detachment rates using the mean field approximation. The analytical results matched very well the Monte Carlo simulations. The main disadvantage of the total lattice model is its inability to account for strong localized effects. We compared our theory with experimental results obtained for ionic self-assembly of silica nanoparticles and found that it could be applied, but it was not as effective as the other models. The comparison with the experimental data leads us to believe that cooperative effects (such as electrostatic interactions, for example) are, in fact, present, but are not relevant for low particle density. 

While the total lattice cooperative model is likely not the best choice for modeling ionic self-assembly of silica nanoparticles, it is designed to model lattice-wide effects and we expect it to match much better to more suitable physical situations. The total lattice cooperative model can be tailored to other systems that involve particle attachment and detachment or even other two-state systems. It can apply to any lattice structure or dimension and can serve as a starting point for other studies.

\section{Final Conclusions}

In this thesis we presented three stochastic model for particle deposition and evaporation. We demonstrated the process of mapping a stochastic model onto the Ising model. We solved each model for a particle density rate equation using the mean field approximation, and validated the results via Monte Carlo simulations. We demonstrated the process of solving for particle density by generation on a Cayley tree and presented results for a general solution for the number of deposited particles using a master equation and matrix theory.

Both the models and the methods presented are very general, and can be easily extended to different physical systems with interactions between nearest-neighbors. We sketch here the basic steps for such an extension:

1. Choose a two-state physical system with interactions between its components. For example: infected/recovered individuals,  two opinion voters, filled/empty parking spots, etc.

2. Choose a topology: grids, trees, complete graphs, etc.

3. Choose a transition rate that describes the interactions between the components of the system.

Once the transition rate is chosen, there are different ways to study such a model.  For the Ising model mapping, one has to impose the detailed balance condition and find the Ising coupling constants in terms of the parameters chosen for the proposed model. Mean field approximations or Monte Carlo simulations can be used for further study into the time development of the system, or to investigate transition rates not amenable to Ising model mapping. The master equation approach can be used to obtain exact solutions for models with more mathematically tractable interactions between particles.

Our project is at the confluence of nanophysics, biology, chemistry, mathematics, and computer science, and provides a pedagogical path toward understanding the complex dynamics of particle self-assembly  by using the tools of statistical physics. We believe that building simple models based on real-life systems leads to valuable pedagogical lessons. Students can see the ``big picture," the link between various disciplines. They learn new ways of thinking about real-life problems and new mathematical and computational techniques. The involvement of students with different science backgrounds in these projects is particularly useful as it stimulates the dialog between disciplines with different perspectives. 

\section*{Acknowledgments}
Special thanks to the my research advisors Dr. I. Mazilu and Dr. D. A. Mazilu. Thanks as well to the rest of my co-authors on the papers connected to this project: Dr. L. Jonathan Cook, B. M. Simpson, A. M. Seredinski, V. O. Kim, W. E. Banks, and B. K. Pope. This research was supported  by the Washington and Lee University Summer Research Scholars program for undergraduate students, the Johnson Opportunity grant and the Lenfest grant.

\end{document}